\let\myover=\over      
\documentclass[10pt,a4paper]{article}
\usepackage{amssymb,amsmath,epsfig,graphics}
\def\be{\begin{equation}}
\def\ee{\end{equation}}
\def\d{\partial}
\def\l{\left(}
\def\r{\right)}
\def\la{\langle }
\def\ra{\rangle }

\newcommand{\bg}{\begin{gather}}
\newcommand{\eg}{\end{gather}}

\def\half{\frac{1}{2}}

\begin{document}
\let\over=\myover  
\def\half{{1 \over 2}} 
\title{Light Sgoldstino: Precision Measurements versus Collider Searches}
\author{ D.~S.~Gorbunov\thanks{{\bf e-mail}: gorby@ms2.inr.ac.ru}, \\
{\small{\em
Institute for Nuclear Research of the Russian Academy of Sciences, }}\\
{\small{\em
60th October Anniversary prospect 7a, Moscow 117312, Russia
}}} 
\date{}
\maketitle
\begin{abstract} 
  \small We study sensitivity of low-energy experiments to the scale
  of supersymmetry breaking $\sqrt{F}$ in models with light
  sgoldstinos --- superpartners of goldstino.  The limits on
  $\sqrt{F}$ may be obtained from direct and indirect measurements of
  sgoldstino coupling to photons, leptons, mesons and nucleons. There
  are three sources of constraints: ($i$) astrophysics and cosmology;
  ($ii$) precision laboratory experiments at low energies; ($iii$)
  rare decays. We discuss only processes with real sgoldstinos. For
  sgoldstino lighter than a few MeV and superpartner masses of the
  order of electroweak scale, astrophysical and reactor bounds on
  $\sqrt{F}$ are significantly stronger than limits which may be
  reached at future colliders. In models with heavier sgoldstino (up
  to 5~GeV), constraints from flavor conserving decays of mesons are
  complementary to ones coming from collider experiments.  The most
  sensitive probes of sgoldstinos are flavor violating processes,
  provided that flavor is violated in squark and/or slepton sector. It
  is shown that sgoldstino contributions into FCNC and lepton flavor
  violation are strong enough to probe the supersymmetry breaking
  scale up to $\sqrt{F}\sim10^7$~GeV, if off-diagonal entries in squark
  (slepton) mass matrices are close to the current limits in MSSM.
\end{abstract} 

\section{Introduction}

Superpartners of goldstino --- longitudinal component of
gravitino --- may be fairly light. In a variety of models with low
energy supersymmetry they are lighter than a few GeV. Such pattern 
emerges in a number of non-minimal 
supergravity models~\cite{ellis,no-scale} and also in gauge mediation
models if supersymmetry is broken via non-trivial superpotential (see,
e.g., Ref.~\cite{gmm} and references therein). 
To understand that superpartners of goldstino may be light, 
it suffices to 
recall that in globally supersymmetric theories with 
canonical K\"ahler potential
and in the absence of anomalous abelian gauge factors, the sum of scalar
squared masses is equal to the sum of fermion squared masses in each
separate sector of the spectrum. Since goldstino is massless,
its spinless superpartners (scalar and
pseudoscalar particles, $S$ and $P$, hereafter, {\it sgoldstinos}) 
are massless too; they are associated with
a non-compact flat direction of the scalar potential. 
Higher order terms from the K\"ahler potential contribute to sgoldstino
masses. Provided these terms are sufficiently suppressed, sgoldstinos
remain light. Of course, these arguments in no way guarantee that
sgoldstinos are always light, but they do indicate that small
sgoldstino masses are rather generic. The theoretical discussion of
sgoldstino masses is contained, e.g., in Ref.~\cite{naturalness}; here
we merely assume that sgoldstinos are light and consider their
phenomenology. 

Sgoldstinos couple to MSSM fields in the same way as goldstino~\cite{bhat}; 
constraints on their couplings may be translated 
into the limits on the supersymmetry breaking parameter $F$. 

There are several papers devoted to 
astrophysical~\cite{9612253}, cosmological~\cite{cosmo-paper} and
collider~\cite{dicus,0001025,0005076} constraints on models with light
sgoldstinos. However, 
the role of light sgoldstinos in low-energy laboratory measurements
has not been studied in detail. To the best of author's knowledge, 
the only paper discussing this issue,
Ref.~\cite{9904367}, concentrated on sgoldstino contribution (as
well as the contribution from light gravitino) into anomalous
magnetic moment of muon. Here we consider a variety of low energy
experiments sensitive to light sgoldstinos. 

In this paper we identify those experiments which are most sensitive to
different sgoldstino vertices for various sgoldstino masses. These
experiments provide constraints on the corresponding coupling
constants. These constants are in fact proportional to the ratios 
of soft terms (squark and gaugino
masses, trilinear coupling constants) and $F$. The latter parameter is
related to the gravitino mass $m_{3/2}$ in a simple way,
$F=\sqrt{3/(8\pi)}m_{3/2}M_{pl}$; small $F$ corresponds to light gravitino
($m_{3/2}<M_{SUSY}$). Hence, the constraints derived in this paper are
of importance for models with light gravitino, whereas sgoldstino
effectively decouple from the visible sector in models with heavy
gravitino. 

In principle, there are
both flavor-conserving and flavor-violating sgoldstino couplings to
fermions. We
present our results in the form of bounds on $\sqrt{F}$ setting soft
flavor-conserving 
terms to be of the order of electroweak scale, as motivated by the 
supersymmetric solution to the gauge hierarchy problem. Flavor-violating
couplings are governed by soft off-diagonal entries in squark
(slepton) squared mass matrices. When evaluating bounds on $\sqrt{F}$ we
set these off-diagonal entries equal to their current limits derived 
from the absence of FCNC
and lepton flavor violation in MSSM~\cite{masiero}. In this way we
estimate the sensitivity of various experiments to the supersymmetry breaking
scale. 

We consider only low-energy processes with sgoldstinos on
mass-shell. Processes with sgoldstino exchange deserve separate
discussion, though we do not expect that the results obtained in this 
paper will be altered significantly. Also, behind the scope of
this paper are loop processes with virtual
sgoldstinos running in loops (for instance,
$K^0-\bar{K}^0$-mixing, $\mu\to e\gamma$, etc.). These processes were analyzed in
Ref.~\cite{gold-flavor} in models with heavy sgoldstinos. Constraints
on $F$ obtained in Ref.~\cite{gold-flavor} are significantly weaker than
ones presented in our paper, so the loop processes are less sensitive
to $F$ in models with heavy sgoldstinos. 
However, models with light sgoldstinos have not been analyzed in
detail yet,
though it was pointed out in Ref.~\cite{gold-flavor} that enhancement
effects may appear if sgoldstinos are light. 
In view of the results obtained in this paper we also find it 
conceivable that light
virtual sgoldstinos may give significant contributions into rare
processes considered in Ref.~\cite{gold-flavor}. 

Let us briefly review the current status of experimental limits on $F$. 
If one ignores sgoldstino, then 
in models with light gravitino the strongest 
direct current bound on $F$ is
obtained from Tevatron, $\sqrt{F}>217$~GeV~\cite{tevatron}. 
In models with light sgoldstinos, collider
experiments become more sensitive to the scale of supersymmetry breaking. 
Namely, LEP and
Tevatron provide constraints at the level of 1~TeV on the 
supersymmetry breaking
scale in models with $m_{S(P)}$ of order of 20~GeV~\cite{0001025,0005076}. 
The most stringent cosmological constraint comes from 
Big Bang Nucleosynthesis~\cite{cosmo-paper}: models with 
light gravitino, $m_{3/2}<1$~eV, that corresponds to
$\sqrt{F}<7\cdot10^4$~GeV, are disfavored if sgoldstinos decouple at 
temperature not less than ${\cal O}(100)$~MeV
($m_{S(P)}\lesssim1$~MeV). It has been argued in Ref.~\cite{9612253} that 
among the astrophysical
constraints, the strongest one comes from SN1987A: 
the gravitino mass is excluded in the range 
$10^{-1.5}~{\rm eV}<m_{3/2}<30$~eV for 1~keV$<m_{S(P)}<10$~MeV and in
a wider range 
$3\cdot10^{-6}~{\rm eV}<m_{3/2}<50$~eV for $m_{S(P)}<1$~keV. 
These excluded intervals correspond to 
$10^4~{\rm GeV}<\sqrt{F}<4\cdot10^5$~GeV 
and $120~{\rm GeV}<\sqrt{F}<5\cdot10^5$~GeV, respectively. 

In this paper we consider various constraints on couplings of light
($m_{S(P)}\lesssim5$~GeV) 
(pseudo)scalars to SM fields coming mostly from astrophysics and
direct precision measurements. So, we partially fill the gap between
constraints coming from collider experiments and cosmology. 

As there are flavor-conserving and flavor-violating interactions of
sgoldstino fields, we have to consider both 
flavor-symmetric and flavor asymmetric
processes. Let us outline our results referring to these two cases in turn. 

We begin with constraints independent of assumptions concerning
breaking of flavor symmetry. 
As expected, strongest bounds arise from astrophysics and
cosmology, that is $\sqrt{F}\gtrsim 10^6$~GeV, or $m_{3/2}>600$~eV, 
for models with $m_{S(P)}<10$~keV and MSSM soft flavor-conserving 
terms being of the order
of electroweak scale. 
For the intermediate sgoldstino masses (up to a few MeV)
constraints from the study of SN explosion and reactor
experiments lead to $\sqrt{F}\gtrsim300$~TeV. 
We will find that for heavier sgoldstinos, low energy processes 
(such as rare decays of mesons) provide 
limits comparable to ones from colliders but valid for different
sgoldstino masses. 

As concerns flavor-asymmetric processes, we find that these 
are generally very sensitive to
light sgoldstino. Namely, with flavor-changing
off-diagonal entries in squark (slepton) squared mass matrix close to
the current bounds, direct
measurements of decays of mesons (leptons) provide very strong bounds,
up to $\sqrt{F}\gtrsim900(15000)$~TeV (valid at $m_{S}\lesssim5(0.34)$~GeV), 
which is much higher than bounds expected from future
colliders. If off-diagonal entries are small, the limits on
$\sqrt{F}$ become 
weaker: they scale as square root of the corresponding off-diagonal elements.

We will see that the rates of processes with one sgoldstino in
final state are proportional to $F^{-2}$, whereas 
the rates of processes with two sgoldstinos in
final state are proportional to $F^{-4}$. 
Hence, under similar assumptions about
soft terms governing sgoldstino couplings, processes with
one sgoldstino are more sensitive to the supersymmetry breaking
scale. Nevertheless, the coupling constants entering one-sgoldstino
and two-sgoldstino processes are generally determined by
different parameters, so the study of two-sgoldstino processes is also
important. 

Further progress in the search for sgoldstino is expected in several
directions. Among the laboratory experiments, the most sensitive to
flavor-conserving sgoldstino coupling for sgoldstino lighter than a
few MeV are experiments with laser photons propagating in magnetic fields and
reactor experiments. For heavier sgoldstinos, measurements of $\Upsilon$
partial widths exhibit the best discovery potential. If flavor
violation in MSSM is sufficiently strong (say, at the level of
current limits), the most promising is the study of charged 
kaon decays. 

This paper is organized as follows. 
In section 2 the effective lagrangian
for sgoldstinos is presented and sgoldstino decay modes are
described. In section 3 we derive various constraints on
the parameter of supersymmetry breaking $\sqrt{F}$ by considering 
low energy processes. 
There we study separately processes with one and
two sgoldstinos in final states (sections 3.1 and 3.2, respectively). 
First, we discuss astrophysical and
cosmological limits on sgoldstino interactions (section 3.1.1). 
Then we present laboratory bounds coming from search
for light (pseudo)scalars in electromagnetic and strong 
processes (section 3.1.2). In sections 3.1.3 and 3.1.4. 
we discuss rare decays with one sgoldstino in final state 
due to flavor-conserving and flavor-violating 
sgoldstino couplings to SM fermions, respectively. Sections 
3.2.1 and 3.2.2 are devoted to rare meson 
decays with two sgoldstinos in final state. 
Our conclusions and comparison
of the results with ones coming from collider experiments are
presented in section 4. 

\section{Effective lagrangian}
\label{sect2}

Let us introduce the effective lagrangian for light goldstino
supermultiplet: scalar $S$, pseudoscalar $P$ and goldstino
$\tilde{G}$. The free part reads
\begin{equation}
{\cal L}=\half\l\d_\mu S\d^\mu S-m_S^2S^2\r+
\half\l\d_\mu P\d^\mu P-m_P^2P^2\r+
{i\over 2}\bar{\tilde{G}}\gamma^\mu\d_\mu\tilde{G}\;.
\nonumber
\end{equation}
There exist two types of interactions in the low-energy effective
theory involving sgoldstino fields: these are terms that couple one
sgoldstino~\cite{bhat,9904367,0001025,0005076} and two 
sgoldstinos~\cite{0001025}, respectively, to SM gauge fields (photons,
gluons) and matter fields (leptons $f_L$, up- and down-quarks $f_U$
and $f_D$). Terms involving one sgoldstino are 
\begin{eqnarray}
{\cal L}_{eff}=-\frac{1}{2\sqrt{2}F}\l m_S^2S\bar{\tilde{G}}\tilde{G}+
im_P^2P\bar{\tilde{G}}\gamma_5\tilde{G}\r\label{eef}
-\frac{1}{2\sqrt{2}}\frac{M_{\gamma\gamma}}{F}SF^{\mu\nu}F_{\mu\nu}+
\frac{1}{4\sqrt{2}}\frac{M_{\gamma\gamma}}{F}
P\epsilon^{\mu\nu\rho\sigma}F_{\mu\nu}F_{\rho\sigma}
-\frac{1}{2\sqrt{2}}\frac{M_3}{F}SG^{\mu\nu~\alpha}G_{\mu\nu}^\alpha\\+
\frac{1}{4\sqrt{2}}\frac{M_3}{F}
P\epsilon^{\mu\nu\rho\sigma}G_{\mu\nu}^\alpha G_{\rho\sigma}^\alpha
-\frac{\tilde{m}_{D_{ij}}^{LR~2}}{\sqrt{2}F}S\bar{f}_{D_i}f_{D_j}-
i\frac{\tilde{m}_{D_{ij}}^{LR~2}}{\sqrt{2}F}P\bar{f}_{D_i}\gamma_5f_{D_j} 
-\frac{\tilde{m}_{U_{ij}}^{LR~2}}{\sqrt{2}F}S\bar{f}_{U_i}f_{U_j}-
i\frac{\tilde{m}_{U_{ij}}^{LR~2}}{\sqrt{2}F}P\bar{f}_{U_i}
\gamma_5f_{U_j}\nonumber\\
-\frac{\tilde{m}_{L_{ij}}^{LR~2}}{\sqrt{2}F}S\bar{f}_{L_i}f_{L_j}-
i\frac{\tilde{m}_{L_{ij}}^{LR~2}}{\sqrt{2}F}P\bar{f}_{L_i}\gamma_5f_{L_j}\;. 
\nonumber
\end{eqnarray}
The direct coupling of two sgoldstinos is described by 
\begin{eqnarray}
{\cal L}_{eff}=\frac{1}{4F^2}\l S\d_\mu P-P\d_\mu S\r
(
(\tilde{m}_{L_{ij}}^{LL~2}+\tilde{m}_{L_{ij}}^{RR~2})\bar{f}_{L_i}
\gamma^\mu\gamma_5f_{L_j}+
(\tilde{m}_{L_{ij}}^{LL~2}-\tilde{m}_{L_{ij}}^{RR~2})\bar{f}_{L_i}
\gamma^\mu f_{L_j}\label{llrrmix}\\+
(\tilde{m}_{D_{ij}}^{LL~2}+\tilde{m}_{D_{ij}}^{RR~2})\bar{f}_{D_i}
\gamma^\mu\gamma_5f_{D_j}
+(\tilde{m}_{D_{ij}}^{LL~2}-\tilde{m}_{D_{ij}}^{RR~2})\bar{f}_{D_i}\gamma^\mu
f_{D_j}
\nonumber\\
+(\tilde{m}_{U_{ij}}^{LL~2}+\tilde{m}_{U_{ij}}^{RR~2})\bar{f}_{U_i}
\gamma^\mu\gamma_5f_{U_j}+
(\tilde{m}_{U_{ij}}^{LL~2}-\tilde{m}_{U_{ij}}^{RR~2})\bar{f}_{U_i}\gamma^\mu
f_{U_j})\;. 
\nonumber
\end{eqnarray}
Here $M_{\gamma\gamma}=M_1\cos^2\theta_W+M_2\sin^2\theta_W$ and $M_i$
are gaugino masses; for 
down-quarks $i=d,s,b$, whereas for up-quarks
$i=u,c,t$; $\tilde{m}_{ij}^{LR~2}$, $\tilde{m}_{ij}^{LL~2}$ and 
$\tilde{m}_{ij}^{RR~2}$ are LR-, LL-, and RR-soft mass terms in 
squark squared mass matrix and for convenience 
we take them real. In what follows we do not
discuss neutrino, so the corresponding couplings are omitted. 
Note that in MSSM the flavor-conserving one-sgoldstino coupling
constants satisfy 
$\tilde{m}_{ii}^{LR~2}=m_{f_i}A_{f_i}$, where $m_{f_i}$ are 
fermion masses and $A_{f_i}$ are 
corresponding soft trilinear coupling constants. Off-diagonal soft
terms $\tilde{m}_{ij}^{LR~2}$, 
$\tilde{m}_{ij}^{LL~2}$ and $\tilde{m}_{ij}^{RR~2}$ are subject to 
constraints from the absence of FCNC and lepton flavor 
violation (see, e.g., Ref.~\cite{masiero}). 

The first part of the effective lagrangian, Eq.~(\ref{eef}), 
is suppressed by $F^{-1}$, whereas the second one,
Eq.~(\ref{llrrmix}), is proportional to $F^{-2}$, 
so processes with two sgoldstinos are very rare. The most 
stringent bounds on $F$ come from processes with one sgoldstino in
final state. 
Nevertheless, as we will see, the absence of processes with two
sgoldstinos gives rise to constraints on supersymmetry breaking
parameter $F$ comparable to bounds from high-energy experiments. The
latter constraints are, strictly speaking, independent of the
constraints coming from one-sgoldstino processes: one-sgoldstino and
direct two-sgoldstino processes are governed by $\tilde{m}^{LR~2}$ and
$\tilde{m}^{LL~2}$, $\tilde{m}^{RR~2}$, respectively.  

Let us discuss decay modes of light sgoldstino. First, sgoldstino decay into
two photons is always open~\cite{0001025}, 
\begin{equation}
\Gamma(S(P)\to\gamma\gamma)=\frac{m_{S(P)}^3M_{\gamma\gamma}^2}{32\pi F^2}\;.
\label{rate-to-photons}
\end{equation}
Second, in models where $m_{3/2}<m_{S(P)}$ sgoldstinos may decay into
gravitino pairs; however, the corresponding 
rates are suppressed by squared ratio of
sgoldstino mass $m_{S(P)}$ and $M_{\gamma\gamma}$ 
in comparison with the decay into two photons, hence this mode may be
disregarded. Third, relatively heavy sgoldstinos
($m_{S(P)}\gtrsim\Lambda_{QCD}$) decay 
into gluons (light mesons) with larger width
than into photons because of color enhancement and because the corresponding
coupling is proportional to gluino mass which is usually the largest
among the gaugino masses, i.e. $M_3>M_{\gamma\gamma}$. 
When analyzing hadronic decay modes of 
light sgoldstinos ($m_{S(P)}<a~few$~GeV),
corresponding rates into quarks and gluons should be rewritten in
terms of light mesons. This step will be presented below. 
Fourth, sgoldstino can decay also
into light leptons if this process is allowed kinematically 
($m_{S(P)}>2m_l$). Since the 
corresponding coupling constants are proportional to
fermion masses these rates are 
suppressed by a factor $m_l^2/m_{P(S)}^2$ apart from the phase space
volume~\cite{0005076}, 
\begin{equation}
\Gamma(S\to l\bar{l})={m_S^3A_l^2\over 16\pi F^2}{m_l^2\over m_S^2}\l
1-{4m_l^2\over m_S^2}\r^{3/2}\;,~~
\Gamma(P\to l\bar{l})={m_P^3A_l^2\over 16\pi F^2}{m_l^2\over m_P^2}\l
1-{4m_l^2\over m_P^2}\r^{1/2}\;.
\label{rate-to-leptons} 
\end{equation}

Consequently, depending on MSSM mass spectrum,
sgoldstino masses and the value of the supersymmetry breaking parameter $F$, 
there are three possible situations in experiments where light
(pseudo)scalar particle appears. This particle may live long enough to escape
from a detector. For instance, in the theory with the superpartner scale of
order 100~GeV and $\sqrt{F}=1$~TeV this behavior would be exhibited by
(pseudo)scalar particle with mass less than 10 MeV, at which sgoldstino 
width is saturated by two-photon mode. Another case is 
when (pseudo)scalar particle decays within detector into two photons or
leptons. Apart from these cases, there is also a possibility of the decay
into two gluons (quarks). For relatively light sgoldstinos (but 
with masses exceeding 270~MeV), the dominant hadronic decay is 
into two pions, while for heavier sgoldstinos $KK$ and
$\eta\eta$ channels become available. Furthermore, there would be 
effects emerging due to $P-\pi^0(\eta,K^0)$ mixing. 

Let us estimate branching ratios of 
hadronic and photonic decay channels neglecting threshold 
factor. In order to estimate sgoldstino coupling to hadrons we
make use of chiral theory of light hadrons. 
There are two different sources of
sgoldstino-meson couplings in the effective lagrangian~(\ref{eef}):
interaction terms with gluons and coupling to quarks. We evaluate
contributions from these two sources into meson-sgoldstino interactions
separately. 

First, we have to relate 
gluonic operators entering Eq.~(\ref{eef}) to meson fields. 
We make use of the correspondence  
\begin{equation}
-\la(\pi\pi)_{J=0}|{\beta(\alpha_s)\over8\pi\alpha_s}
G_{\mu\nu}^aG^{a~\mu\nu}|0\ra=\half
q^2\varphi_\pi^\alpha\varphi_\pi^\alpha
\label{gluons-mesons}
\end{equation}
derived in Ref.~\cite{voloshin-zakharov}. Here $q^2$ is momentum
of pion pair created with zero 
total angular momentum, $J=0$; $\beta(\alpha_s)$ is the $\beta$-function
of QCD, $\varphi_\pi^\alpha$ is 
the pion isotopic amplitude, 
$$
\varphi_\pi^\alpha\varphi_\pi^\alpha=2
\varphi_{\pi^+}\varphi_{\pi^-}+\varphi_{\pi^0}\varphi_{\pi^0}\;,
$$
and quarks and mesons are considered 
massless. At higher energies also $KK$ and
$\eta\eta$ pairs may be created by gluonic operator. 

There is one more 
relation~\cite{novikov},
\begin{equation}
\la A|{N_f\alpha_s\over 4\pi}G_{\mu\nu}^a\tilde{G}^{a~\mu\nu}|0\ra
=const\cdot\epsilon\cdot f_A m_A^2\varphi_A\;,
\label{gluons-pseudoscalar}
\end{equation} 
where $\tilde{G}^{a~\mu\nu}$ is a tensor dual to gluonic one, 
$A$ is a neutral 
pseudoscalar meson ($\pi^0$, $\eta$) and $const$
is a normalization factor; $f_A=f_\pi=130$~MeV and $\epsilon$ is a
parameter responsible for $SU(N_f)$ flavor symmetry breaking
($\epsilon=(m_u-m_d)/(m_u+m_d)$ for $\pi^0$, 
$\epsilon\simeq1$ for $\eta$). 

In fact, the lagrangian~(\ref{eef}) describes sgoldstino interactions at
the superpartner scale. Sgoldstino coupling constants at low energies may
be obtained by making use of renormalization group evolution. Thus for
the gluonic operator one has
\begin{equation}
G_{\mu\nu}^2(M_3)=G_{\mu\nu}^2(\mu){\beta(\alpha_s(\mu))\over\alpha_s(\mu)}
{\alpha_s(M_3)\over\beta(\alpha_s(M_3))}\;.
\nonumber
\end{equation} 
Hence, we estimate the matrix element of the gluonic operator 
between the scalar and meson pair as 
\begin{equation}
\la (AA)_{J=0}|{M_3\over
2\sqrt{2}F}G_{\mu\nu}^aG^{a~\mu\nu}S|S\ra=
{\alpha_s(M_3)\over\beta(\alpha_s(M_3))}
q^2\sqrt{2}\pi \varphi_A\varphi_A
{M_3\over F}\varphi_S
\label{sgoldstino-mesons}
\end{equation} 
and in a similar way we estimate the matrix element of another 
gluonic operator between the pseudoscalar and meson
\begin{equation}
\la A|{M_3\over
2\sqrt{2}F}G_{\mu\nu}^a\tilde{G}^{a~\mu\nu}P|P\ra=
{\epsilon\cdot const\over\alpha_s(M_3)}
{\sqrt{2}\pi\over N_f}f_A m_A^2\varphi_A 
{M_3\over F}\varphi_P\;.
\label{sgoldstino-meson}
\end{equation}  
Note, that these matrix elements are highly suppressed by squared
sgoldstino or meson masses.  

Since direct sgoldstino coupling to 
quarks contributes also to meson production, we remind basic
relations of chiral theory
\begin{equation}
\la 0|J^{\pi^0}_\mu(0)|\pi^0(\vec{q})\ra=\frac{i}{\sqrt{2}}f_\pi
q_\mu\;,~~~
\la 0|J^{\pi^+}_\mu(0)|\pi^+(\vec{q})\ra=if_\pi
q_\mu\;.
\label{pion-decay-constant-1}
\end{equation}
where 
\begin{equation}
J^{\pi^0}_\mu=
\half\l\bar{u}\gamma_\mu\gamma_5u-\bar{d}\gamma_\mu\gamma_5d\r\;,~~~~
J^{\pi^+}_\mu=\bar{d}\gamma_\mu\gamma_5u\;. 
\label{pion-decay-constant-2}
\end{equation}
If we parameterize sgoldstino couplings to the triplet of light quarks $q$ as 
$$
{\cal L}=-\bar{q}\l S\hat{\Sigma}_S-i\gamma_5 P\hat{\Sigma}_P\r q
$$
with $\hat{\Sigma}_S$ and $\hat{\Sigma}_P$ 
being $3\times 3$ matrices of the corresponding
coupling constants (which are read off from Eq.~(\ref{eef})), 
then the 
standard procedure (see, e.g., Ref.~\cite{pich}) gives the following
low-energy effective lagrangian
\begin{equation}
{\cal L}_{meson}=B_0{\rm Tr}\l
f_{\pi^0}\hat{\Phi}\hat{\Sigma}_P P-S\hat{\Sigma}_S\hat{\Phi}^2\r
\label{meson-sgoldstino-eff-lagr}
\end{equation}
to the leading order in mesonic fields included in 
matrix $\hat{\Phi}$. The constant $B_0$ is related to
quark condensate as $\la 0|\bar{q}q|0\ra=-\half B_0f_{\pi^0}^2$ and may be
evaluated from the masses of kaon and quarks,
$B_0=M_{K^0}^2/(m_d+m_s)$. We 
account only for one-sgoldstino terms since others are suppressed by
sgoldstino masses and additional inverse power of $F$. 

The lagrangian~(\ref{meson-sgoldstino-eff-lagr}) 
consists of two parts. The first one, 
\begin{equation}
{\cal L}_{meson-1}=-\frac{B_0f_{\pi^0}}{\sqrt{2}F}\l\frac{\pi^0}{\sqrt{2}}
\l
\tilde{m}_{U_{11}}^{LR~2}-\tilde{m}_{D_{11}}^{LR~2}\r+\frac{\eta}{\sqrt{6}}
\l\tilde{m}_{U_{11}}^{LR~2}+\tilde{m}_{D_{11}}^{LR~2}-
2\tilde{m}_{D_{22}}^{LR~2}\r+K^0\tilde{m}_{D_{21}}^{LR~2}
+\bar{K^0}\tilde{m}_{D_{12}}^{LR~2}\r P\;,
\label{first-part}
\end{equation}
is pseudoscalar sgoldstino mixing with $\pi^0$, $\eta$, $K^0$ and $\bar{K^0}$
mesons, while the second one,
\begin{eqnarray}
{\cal L}_{meson-2}=-\frac{B_0}{\sqrt{2}F}
\Biggl(\l\tilde{m}_{U_{11}}^{LR~2}+\tilde{m}_{D_{11}}^{LR~2}\r
\l\pi^+\pi^-+K^+K^-+K^0\bar{K^0}+\half\pi^0\pi^0\r\label{second-part}
\\\nonumber
+\frac{1}{6}\eta^2\l\tilde{m}_{U_{11}}^{LR~2}+
\tilde{m}_{D_{11}}^{LR~2}+
4\tilde{m}_{D_{22}}^{LR~2}\r+\frac{1}{\sqrt{3}}\pi^0\eta
\l\tilde{m}_{U_{11}}^{LR~2}-\tilde{m}_{D_{11}}^{LR~2}\r-
\frac{1}{\sqrt{2}}\pi^0\bar{K^0}\tilde{m}_{D_{12}}^{LR~2}-
\frac{1}{\sqrt{2}}\pi^0K^0\tilde{m}_{D_{21}}^{LR~2}\\\nonumber
-\frac{1}{\sqrt{6}}\bar{K^0}\eta\tilde{m}_{D_{12}}^{LR~2}
-\frac{1}{\sqrt{6}}K^0\eta\tilde{m}_{D_{21}}^{LR~2}+K^-
\pi^+\tilde{m}_{D_{12}}^{LR~2}+K^+
\pi^-\tilde{m}_{D_{21}}^{LR~2}\Biggr)S
\end{eqnarray}   
describes scalar sgoldstino decays into mesons.  
Note that sgoldstino couplings with two different mesons is
suppressed by off-diagonal term in squark mass matrix. In what follows
we will not consider processes where real sgoldstino decays into such 
modes. 

Now let us estimate matrix elements between sgoldstino and
meson (i.e., sgoldstino-meson mixing) 
as a sum of two quantities, Eq.~(\ref{sgoldstino-meson}) and
Eq.~(\ref{first-part}), while the amplitude of the scalar 
sgoldstino decay into
pairs of light mesons is evaluated as a sum 
of Eq.~(\ref{sgoldstino-mesons}) and
Eq.~(\ref{second-part}). Let us 
compare contributions of gluon and quark operators into
sgoldstino couplings to mesons. As an example, for the ratio of
the corresponding contributions into coupling of the scalar 
to neutral pions and into pion-pseudoscalar mixing we obtain
$$
\frac{\la\pi^0\pi^0|S\ra_{gluon}}{\la\pi^0\pi^0|S\ra_{quark}}
={\alpha_s(M_3)\over\beta(\alpha_s(M_3))}
4\pi{m_S\over B_0}{M_3\over A_Q}{m_S\over
m_u+m_d}\;,
$$
$$
\frac{\la\pi^0|P\ra_{gluon}}{\la\pi^0|P\ra_{quark}}
={2\pi\sqrt{2}\over\alpha_s(M_3)}{M_3\over 3
A_Q}\;.
$$
These ratios 
are larger than 10 for $M_3=A_Q$. Hence, gluonic
operators give rise to stronger coupling 
of light sgoldstinos to light 
mesons, as compared to sgoldstino-quark interactions. 

Let us evaluate the rate of the scalar sgoldstino decay into light mesons,
assuming that 
this decay is allowed kinematically. As an example, for the neutral
pion mode we obtain
\begin{equation}
\Gamma(S\to\pi^0\pi^0)={\alpha^2_s(M_3)\over\beta^2(\alpha_s(M_3))}
{\pi m_S\over
324}{m_S^2M_3^2\over F^2}\l1-{\beta(\alpha_s(M_3))\over\alpha_s(M_3)}
{9\over 4\pi}{B_0\over
m_S}{m_u+m_d\over m_S}{A_Q\over M_3}\r^2\sqrt{1-{4m_{\pi^0}^2\over m_S^2}}\;.
\nonumber
\end{equation} 
Taking into account only the largest contribution from the gluon operator 
and neglecting the threshold factor 
we estimate the ratio of rates of sgoldstino
decays into photons and mesons,
\begin{equation}
{\Gamma(S\to\gamma\gamma)\over\Gamma(S\to\pi^0\pi^0)}={81\over8\pi^2}
{\beta^2(\alpha_s(M_3))\over\alpha^2_s(M_3)}{M_{\gamma\gamma}^2\over M_3^2}\;.
\nonumber
\end{equation} 
We see that this ratio is smaller than 1 at 
$M_{\gamma\gamma}=M_3$. Since in most models gluino is
several times heavier than photino, for sufficiently heavy
sgoldstinos hadronic modes usually dominate over photonic one.   

Let us estimate now the contribution of pion-sgoldstino mixing
into pseudoscalar sgoldstino width. 
Recall that the pion width is almost saturated by
the two-photon decay mode. Then 
\begin{equation}
\Gamma(P\to \pi^0\to
\gamma\gamma)={1\over\alpha^2_s(M_3)}
{\pi^2f_{\pi^0}^2\over 4\l
m_P^2-m_{\pi^0}^2\r^2}
{M_3^2m_{\pi^0}^4\over F^2}\l{m_u-m_d\over m_u+m_d}\r^2
\Gamma^*(\pi^0\to\gamma\gamma)\;,
\label{pion-goldstino-mixing}
\end{equation}
where the 
two-photon width of virtual pion is taken at $p_{\pi^0}^2=m_P^2$ and may be
approximated as 
$$
\Gamma^*(\pi^0\to\gamma\gamma)\approx\Gamma_{tot}(\pi^0){m_P^3\over m_{\pi^0}^3}\;.
$$
With account of only 
leading contributions from gluonic operator we obtain 
\begin{equation}
{\Gamma^{direct}(P\to\gamma\gamma)\over\Gamma(P\to \pi^0\to\gamma\gamma)}=
\frac{\alpha^2_s(M_3)}{8\pi^3}\tau_{\pi^0}{m_{\pi^0}}{M_{\gamma\gamma}^2\over
M_3^2}{m_{\pi^0}^2\over f_{\pi^0}^2}\l{m_P^2\over m_{\pi^0}^2}-1\r^2\;.
\label{branchings-2}
\end{equation}
As discussed above, $\tilde{m}_{ii}^{LR~2}=m_iA_{Q_i}$, so at
$M_{\gamma\gamma}=M_3$ and light $P$ ($m_P\ll m_{\pi^0}$) 
we obtain that the ratio~(\ref{branchings-2}) is numerically
$8\cdot10^2$. In the opposite case of heavy $P$ 
($m_P\gg m_{\pi^0}$) the ratio becomes even larger. 
Hence mixing with pions gives negligible contribution to 
sgoldstino decay into photons (unless $M_{\gamma\gamma}\lesssim
M_3/30$; we do not consider this case). 
The only exception is the degenerate case, when sgoldstino and pion
masses are close and this branching becomes of order
1. (In the case of strong degeneracy there is also a correction to pion
life-time which may give rise to a constraint on $F$). We
do not consider this unrealistic situation. The interference with 
$\eta$-meson gives nothing new. Indeed,
Eq.~(\ref{branchings-2}) scales as $\tau_{meson}m_{meson}^3$
which is invariant
under the variation of meson mass, if the meson 
width is (almost) saturated by
anomalous decay into two photons. Decay via neutral kaon is also
negligible because of large kaon life-time.  

\begin{figure}[htb]
\begin{center}
{\epsfig{file=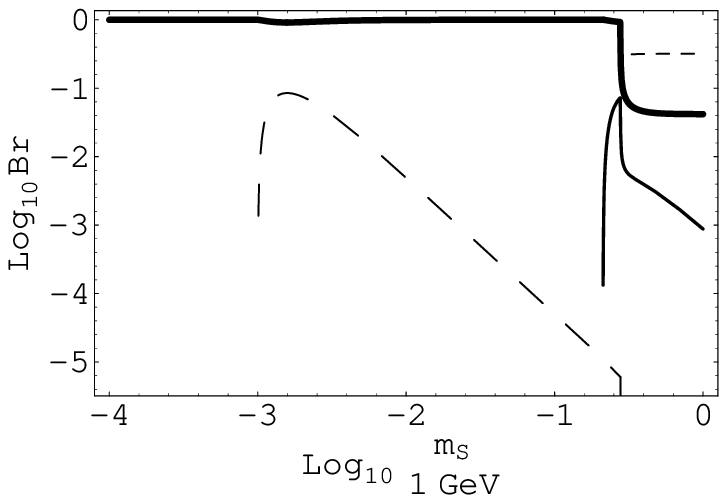,height=5cm,width=7.5cm}}
{\epsfig{file=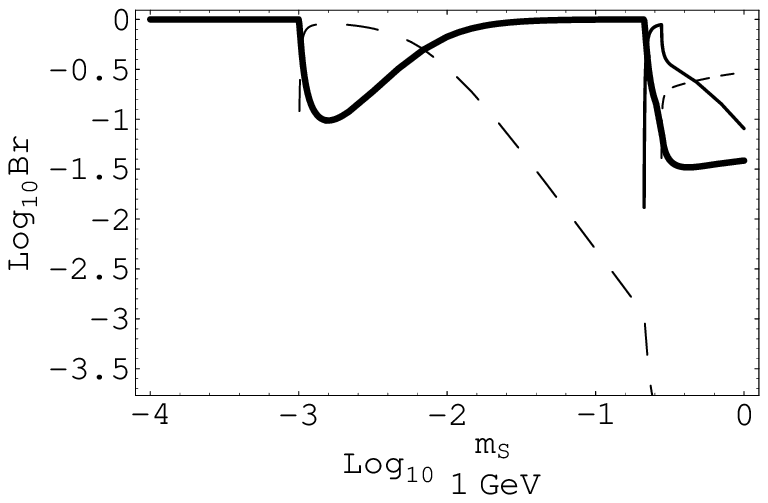,height=5cm,width=7.5cm}}
{\epsfig{file=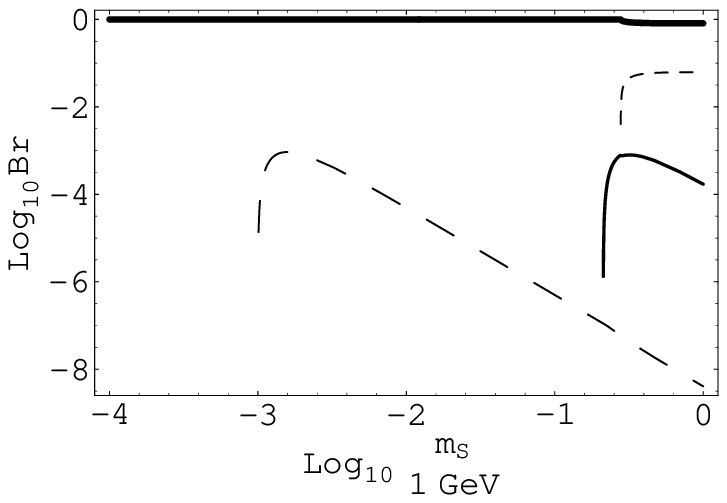,height=5cm,width=7.5cm}}
{\epsfig{file=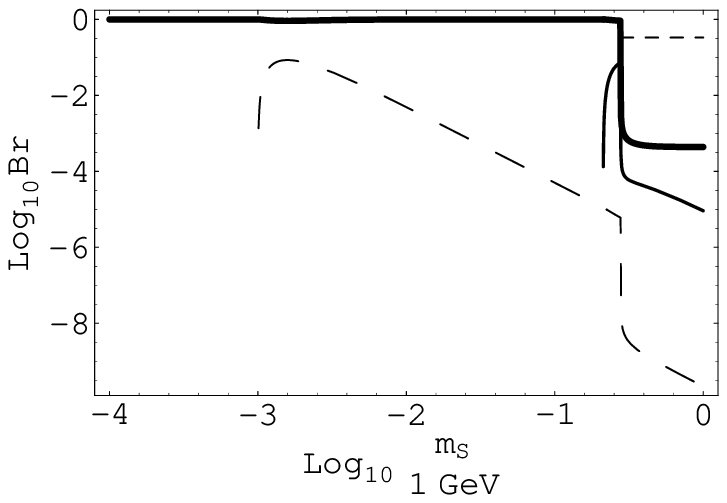,height=5cm,width=7.5cm}}
\caption{Branching ratios of scalar sgoldstino decays into photons
(thick line), $e^+e^-$ (long dashed line), $\mu^+\mu^-$ (thin line)
and $\pi^0\pi^0$ (short dashed line) in models with: a)
$|A|=M_{\gamma\gamma}=M_3=100$~GeV; b) $M_{\gamma\gamma}=M_3=100$~GeV,
$|A|=1$~TeV; c) $A=M_3=100$~GeV, $M_{\gamma\gamma}=1$~TeV; d)
 $|A|=M_{\gamma\gamma}=100$~GeV, $M_3=1$~TeV.}
\label{Jfig}
\end{center}
\end{figure}

To conclude this section we summarize the situation with 
sgoldstino branching ratios. Let us begin with scalar sgoldstino. 
In Figure~\ref{Jfig}
\begin{figure}[htb]
\begin{picture}(0,0)(-38,86)
\put(130.00,177.00){\line(1,0){24.00}}  
\put(20.00,310.00){\makebox(0,0)[cb]{c)}}
\put(130.00,321.00){\line(1,0){24.00}}  
\put(20.00,450.00){\makebox(0,0)[cb]{a)}} 
\put(350.00,177.00){\line(1,0){24.00}}  
\put(240.00,310.00){\makebox(0,0)[cb]{d)}}
\put(354.00,320.00){\line(1,0){24.00}}  
\put(240.00,450.00){\makebox(0,0)[cb]{b)}} 
\end{picture}
\end{figure}
we present 
scalar sgoldstino branching ratios into photons, leptons and
neutral pions evaluated for four different sets of supersymmetry
breaking soft terms, $A$, $M_{\gamma\gamma}$, $M_3$. To determine
photonic and leptonic widths we make use of 
Eqs.~(\ref{rate-to-photons}) and (\ref{rate-to-leptons}), while
the width into two neutral pions is calculated according to 
Eq.~(\ref{gluons-mesons}) generalized to 
non-zero pion masses. Estimating hadronic sgoldstino partial 
width we account
only for $\pi^+\pi^-$ and $\pi^0\pi^0$ decay modes. Other hadronic
modes may be considered in the same way. 
Ratios between different hadronic channels are determined by
chiral theory.   

Scalar sgoldstino lighter than
270~MeV almost always 
predominantly decays into two photons. At sgoldstino mass close to $2m_e$ or
$2m_\mu$, rates of the decays into pairs of 
corresponding leptons become comparable to
the two-photon rate and even exceed the latter 
in models with large trilinear soft 
terms. Far from the lepton mass, the corresponding lepton branching
ratio decreases as $m_l^2/m_{S(P)}^2$. At sgoldstino masses exceeding
270~MeV hadronic modes emerge. Their rates are somewhat higher than
the rate of the two-photon decay except for models with large
$M_{\gamma\gamma}$, in which the photonic mode always dominates. 

As regards pseudoscalar sgoldstino, it does not have the decay mode 
into two pseudoscalar mesons to the zero order in $G_F$. 
Hence at $M_{\gamma\gamma}\sim M_3$ its
hadronic decay modes are suppressed unless $m_P$ is quite large (well
above 1~GeV). In what follows we consider photonic and leptonic decay
channels of the pseudoscalar sgoldstino only.  

\section{Searches for light sgoldstino}
In accordance with the discussion of sgoldstino effective lagrangian
presented in the previous section, 
there are two types of processes we are interested in. In the
processes of the first type 
only one sgoldstino emerges while in the processes of the second type a pair of
sgoldstinos appears in the final state. These processes are governed
by different coupling constants and will be considered in turn.  

\subsection{Processes with one sgoldstino in the final state}
\subsubsection{Bounds from astrophysics and cosmology}

In subsections 3.1.1 and 3.1.2 
we consider mainly pseudoscalar sgoldstino, though
almost all constraints are valid for the scalar sgoldstino 
as well; a few exceptions will be pointed out.  

Light pseudoscalar particles appear in particle physics models in
various contexts; a well known example is an axion. There are numerous
cosmological, astrophysical and laboratory bounds on interactions of
light pseudoscalars which apply to sgoldstino. For completeness we
collect in sections 3.1.1 and 3.1.2 the most stringent of these bounds
and translate the bounds on sgoldstino coupling constants into bounds
on supersymmetry breaking parameters $\sqrt{F}$ and $m_{3/2}$. Let us
write the interactions of sgoldstino with photons and fermions as
follows 
\begin{equation}
{\cal L}_{\gamma P}=-\sqrt{2}{M_{\gamma\gamma}\over
F}P\vec{E}\vec{B}\equiv g_{\gamma}P\vec{E}\vec{B}\;,~~~
{\cal L}_{fP}={m_fA_f\over \sqrt{2}F}P\bar{f}\gamma_5f 
\equiv g_{fP}P\bar{f}\gamma_5f\;. 
\label{aaa}
\end{equation}
Then the limits on $g_\gamma$ and $g_{fP}$ imply limits on
$\sqrt{F}$. 

In what follows we set 
$M_{\gamma\gamma}=A_f=100$~GeV in our quantitative estimates, since
superpartner scale $M_{SUSY}$ is expected to be close to
the electroweak scale as motivated by supersymmetric solution to the hierarchy
problem in SM. 

One of the sources of pseudoscalars are stars: light pseudoscalars are
produced there by Primakoff process, that is $\gamma\to P$ conversion
in external electromagnetic field. 
Another place of sgoldstino creation is galactic space where
magnetic fields produce pseudoscalars from propagating
photons. 
\begin{table}[htb]
$$
\begin{array}{|c|c|c|c|c|}
\hline {\rm Experiment} & m_P & g_\gamma\;,{\rm GeV}^{-1} & \sqrt{F}\;,{\rm GeV} & m_{3/2} \\
\hline
``helioscope''~\cite{HelScope} & \lesssim0.03~{\rm eV} & 
<6\cdot10^{-10} & >0.5\cdot 10^6 & 
>60~{\rm eV}\\
\hline
SOLAX~\cite{Solar-crystal} & <1~{\rm keV} & <2.7\cdot10^{-9} & 
>2.3\cdot 10^5 & >12~{\rm eV}\\
\hline       
SN~\cite{SN} & <10^{-9}~{\rm eV} & <10^{-11} & 
>4\cdot 10^6 & >3.5~{\rm keV}\\              
\hline
HBS~\cite{HBS} & <10~{\rm keV} & <6\cdot 10^{-11} & 
>1.5\cdot 10^6 & >550~{\rm eV}\\
\hline
Photon & & <10^{-14} & 
> 1.2\cdot10^8& 
>3.5~{\rm MeV}\\
Background~\cite{Back} & 1~{\rm keV} & {\rm or} &{\rm or} 
 &{\rm or} 
\\
and & & >10^{-5} & 
<4\cdot10^3 & 
<3.5~{\rm meV}\\

distortion~of & & <10^{-20} & 
>1.2\cdot 10^{11} & 
>3.5~{\rm TeV}\\
CMBR& 1~{\rm MeV} &{\rm or}  &{\rm or} 
&{\rm or} 
\\
spectrum~\cite{Back} &&>10^{-9}  & 
<4\cdot10^5& 
<35~{\rm eV}\\
\hline
& & <5.6\cdot10^{-10} & 
>5\cdot 10^5 & 
>50~{\rm eV}\\
SN1987A~\cite{9612253} & <1~{\rm keV} & {\rm or} & 
{\rm or} & 
{\rm or}\\
& & >10^{-2} & 
<120 & 
<3\cdot10^{-6}~{\rm eV}\\
\hline
& & <9\cdot10^{-10} & 
>4\cdot 10^5 & 
>30~{\rm eV}\\
SN1987A~\cite{9612253} & 1~{\rm keV}<m_P<10~{\rm MeV} & {\rm or} & 
{\rm or} & 
{\rm or}\\
& & >8\cdot10^{-7} & 
<1.3\cdot10^4 & 
<10^{-1.5}~{\rm eV}\\
\hline
\end{array}
$$
\caption{Constraints from astrophysics and cosmology on SUSY models
with light sgoldstinos coupled to photons.}
\label{tab-astro-sgold-photon}
\end{table}

In ``helioscope'' method, a dipole magnet directed towards the
Sun is used. Inside the volume with strong magnetic field, 
solar pseudoscalars can transform into X-rays by inverse Primakoff
process. An alternative method, ``Bragg diffraction'', was applied in
SOLAX experiment to detect solar pseudoscalars. The absence of
anomalous X-ray fluxes from SN1987A related to possible pseudoscalar
conversion into photons in galactic magnetic field gives the
strongest constraint on $g_\gamma$. Since this limit is valid only for
unrealistically light pseudoscalar ($m_P<10^{-9}$~eV), we consider the 
helium-burning life-time of Horizontal Branch
Stars (HBS) in globular clusters as the most sensitive probe of $F$ at
very small $m_P$. 

There are two more constraints on $g_\gamma$ coming from cosmology and
astrophysics. Light sgoldstinos are thermally produced in the early
Universe via Compton process $e\gamma\to eP(S)$. Photons from 
sgoldstino decays contribute to the photon extragalactic background,
if sgoldstinos outlive matter-radiation decoupling. If, on the other
hand, sgoldstinos
decay before matter-radiation decoupling, produced photons may heat
electrons leading to distortion of 
CMBR spectrum, which is experimentally studied well enough to exclude wide
range of $F$ at
corresponding sgoldstino masses. The experiments on
photon background and cosmic microwave background radiation, being
combined, exclude a
strip in $(m_P,g_\gamma)$ plane (see Ref.~\cite{Back}). 
In Table~\ref{tab-astro-sgold-photon} we present the corresponding
limits for two typical values of $m_P$.

All these
constraints~\footnote{See also Ref.~\cite{Back} for 
constraints on $g_\gamma$ coming from 
Deuterium fission by scalars decaying into two photons.} 
on $g_\gamma$ are collected in 
Table~\ref{tab-astro-sgold-photon}. The 
limits on $\sqrt{F}$ 
are obtained at $M_{\gamma\gamma}=100$~GeV and scale as square
root of $M_{\gamma\gamma}$. For completeness, we included in 
Table~\ref{tab-astro-sgold-photon} also the limits obtained in 
Ref.~\cite{9612253} by considering SN1987A. 
 
\begin{table}[htb]
$$
\begin{array}{|c|c|c|c|c|}
\hline {\rm Experiment} & m_P & g_f & \sqrt{F}\;,{\rm GeV} & m_{3/2} \\
\hline Red~Giants~\cite{RedGiant} & <10~{\rm keV} &
g_e<2.5\cdot10^{-13} & >3.8\cdot 10^5 & >35~{\rm eV}\\ 
\hline
HBS~\cite{HBS-eP,RedGiant} & \lesssim 10~{\rm keV} &
g_{eP}<0.5\cdot10^{-12} & >2.7\cdot 10^5 & >17~{\rm eV} \\ 
\hline
HBS~\cite{HBS-eS} & \lesssim 10~{\rm keV} &
g_{eS}<1.3\cdot10^{-14} & >1.6\cdot 10^6 & >650~{\rm eV} \\ 
\hline
HBS~\cite{HBS-eS} & \lesssim 1~{\rm keV} &
g^{(0)}_N<4.3\cdot10^{-11} & \gtrsim 1.2\cdot10^6 & >370~{\rm eV} \\ 
\hline &  &
g^{(0)}_N<3\cdot10^{-10} & \gtrsim 5\cdot10^5 & >50~{\rm eV} \\ 
SN1987A~\cite{R} & \lesssim 10~{\rm MeV} &
{\rm or} & {\rm or} & {\rm or} \\ 
 &  &
g^{(0)}_N>3\cdot10^{-7} & \lesssim 1.5\cdot10^4 & <0.05~{\rm eV} \\ 
\hline
\end{array}
$$
\caption{Constraints from Astrophysics on SUSY models
with light sgoldstinos coupled to fermions.}
\label{astro-sgoldstino-fermions}
\end{table}

Let us proceed with sgoldstino coupling to electrons. 
Restrictive limits come from
delay of helium ignition in low-mass red giants. 
There are also two limits
on coupling to electrons 
from bremsstrahlung process $e^-+(A,Z)\to(A,Z)+e^-+P$ and 
Compton process $\gamma+e^-\to e^-+P$ in stars: these processes lead 
to energy loss of stars and are constrained by helium-burning
life-time of Horizontal Branch Stars. Note that the life-time of HBS
gives stronger constraints on electron coupling to scalar than to
pseudoscalar.  

Let us turn to (pseudo)scalar coupling to nucleons. 
In order to relate the corresponding constant $g_N$ to $F$ we make 
use of the analogy to axion. Then effective lagrangian reads
\begin{equation}
{\cal L}_{eff}=i\bar{\psi}\gamma_5(g_N^{(0)}+g_N^{(3)}\tau_3)\psi P\;,
\label{*}
\end{equation}
where $\psi$ denotes the nucleon dublet and 
\begin{equation}
g_N^{(0)}\sim{A_Qm_N\over\sqrt{2}F}\;,
~~~g_N^{(3)}\sim{m_u-m_d\over m_u+m_d}g_N^{(0)}\;.
\label{**}
\end{equation}
The energy loss of Horizontal Branch Stars via 
Compton process $\gamma+^4\!\!{\rm He}\to^4\!\!{\rm He}+S$ 
gives rise to a bound on $F$. Also, nucleon-sgoldstino coupling leads to 
shortening of SN1987A neutrino burst. 

Astrophysical constraints on sgoldstino-fermion interactions are presented in
Table~\ref{astro-sgoldstino-fermions}. Bounds on $\sqrt{F}$ are
obtained at $A_e=A_Q=100$~GeV and scale as $\sqrt{A_f}$, $f=Q,e$. 
Note that the region $\sqrt{F}\lesssim 1.5\cdot10^4$~GeV allowed 
by SN explosion~\cite{R} is not ruled out by astrophysical
arguments or direct measurements if sgoldstino is relatively heavy
(10~keV$\lesssim m_{S(P)}\lesssim$10~MeV) and its interactions conserve
flavor (see below). 

For constraints coming from Big Bang Nucleosynthesis see 
Ref.~\cite{cosmo-paper}. 

\subsubsection{Laboratory bounds on very light sgoldstinos}
Let us now consider direct laboratory limits on couplings of very
light sgoldstinos. 
\begin{table}[htb]
$$
\begin{array}{|c|c|c|c|c|}
\hline {\rm Experiment} & m_P 
& g_\gamma\;,{\rm GeV}^{-1} & \sqrt{F}\;,{\rm GeV} & m_{3/2}\;,{\rm meV} \\
\hline Laser~\cite{cameron} & <10^{-3}~{\rm eV} &
<3.6\cdot10^{-7} & >2.0\cdot 10^4 & >93\\ 
\hline
\gamma-regeneration~\cite{cameron} & <10^{-3}~{\rm eV} &
<6.7\cdot10^{-7} & >1.5\cdot 10^4 & >50 \\ 
\hline
NOMAD~\cite{NOMAD} &\lesssim 40~{\rm eV} &
<1.5\cdot10^{-4} & >970 & >0.22 \\ 
\hline
\end{array}
$$
\caption{Constraints from direct measurements on SUSY models
with light sgoldstinos}
\label{low-sgoldstino-photons}
\end{table}

The first set of bounds on $F$ 
comes from the study of laser
beam propagation through
transverse magnetic field. The production of real sgoldstinos would
induce the rotation of the beam polarization, while the emission and
absorption of virtual sgoldstinos would contribute to the ellipticity
of the laser beam. Such effects have not been observed and their absence
implies a constraint on pseudoscalar-photon coupling. 
There is also a constraint on the interaction of a 
pseudoscalar particle 
with photons coming from experiments on photon regeneration. 
In these experiments, light pseudoscalars
produced via Primakoff effect penetrate through optic
shield and then transform back into photons 
(``invisible light shining through walls''). Similar scheme is applied
in NOMAD experiment. The results are presented in 
Table~\ref{low-sgoldstino-photons} at $M_{\gamma\gamma}=100$~GeV; 
limits on $\sqrt{F}$ scale as $\sqrt{M_{\gamma\gamma}}$ whereas bounds
on $m_{3/2}$ scale as $M_{\gamma\gamma}$.  
\begin{table}[htb]
$$
\begin{array}{|c|c|c|c|c|}
\hline m_P & {\rm Final~state}, X & g_N^{(0)}\cdot{\rm Br}^{1/2}_{(P\to X)} 
& \sqrt{F}\cdot{\rm Br}^{-1/4}_{(P\to X)}\;,{\rm GeV} 
& m_{3/2}\cdot{\rm Br}^{-1/2}_{(P\to X)} \\
\hline <1.5~{\rm MeV} & e^+e^- &
\lesssim7\cdot10^{-10}~\cite{altmann} 
& \gtrsim3\cdot 10^5 
& \gtrsim25~{\rm eV} \\ 
\hline <1~{\rm MeV} &\gamma\gamma
&\lesssim8\cdot10^{-7}~\cite{koch} 
& \gtrsim9\cdot10^3 
& \gtrsim20~{\rm meV} \\ 
\hline
\end{array}
$$
\caption{Constraints from reactor experiments on SUSY models
with light sgoldstinos}
\label{low-sgoldstino-reactor}
\end{table}

Another set of constraints is obtained from reactor
experiments, where nuclear de-excitation is studied. 
Let us again make use of Eq.~(\ref{*}) and Eq.~(\ref{**}). 
Then we obtain for the 
isoscalar transition from excited nucleon state with the change of
spin by $J$ and isospin by $T$ and with emission of photons and
pseudoscalars with momenta $k_\gamma$ and $k_P$, respectively, the
following ratio of rates~\cite{reactor}
$$
{\omega_P^{J=1,T=0}\over\omega_\gamma^{M1,J=0}}\simeq 6
\l{k_P\over k_\gamma}\r^3
{g^{(0)2}g_{\pi NN}^2f_\pi^2\over 4\pi\alpha M_N^2}\;,  
$$
where $M1$ refers to the type of electromagnetic transition and effective
pion-nucleon coupling constant is $g_{\pi NN}^2/4\pi=14.6$. Products of
pseudoscalar decay (two photons or $e^+e^-$) are observed in 
detectors. In this way two constraints on the coupling of a
pseudoscalar to nucleon have been obtained: 
$\omega_P/\omega_\gamma\times
{\rm Br}(P\to e^+e^-)<10^{-16}$~\cite{altmann} and 
$\omega_P/\omega_\gamma\times
{\rm Br}(P\to\gamma\gamma)<1.5\cdot
10^{-10}$~\cite{koch} (we set the pseudoscalar momentum 
equal to photon frequency, $k_P=k_\gamma$). Corresponding bounds on
$\sqrt{F}$ are presented in Table~\ref{low-sgoldstino-reactor} 
at $A_Q=100$~GeV and 
scale as $\sqrt{A_Q}$. The first constraint is
valid for $m_P<1.5$~MeV and becomes weaker for heavier sgoldstinos,
while the second limit is relevant only for light sgoldstino, $m_P<1$~MeV. 
The larger the branching
ratio the stronger the corresponding bounds on $\sqrt{F}$: these bounds
scale as quartic root of branching ratios. Although sgoldstino
branching into $e^+e^-$ is usually very small (see Figure~\ref{Jfig}),
current experimental bounds
on $F$ from sgoldstino decaying into $e^+e^-$ are stronger than
limits from decay into two photons. 
Note that reactor experiments give fairly
strong bounds on $F$ but they should be considered as order-of
magnitude estimates, as obtaining exact numbers requires accurate
calculations involving nuclear matrix elements. 

\subsubsection{Flavor conserving rare decays}
Numerous bounds arise from precise measurements of partial
widths of mesons and leptons (see
Tables~\ref{table-sgoldstino-rare-decays-out}, 
\ref{table-sgoldstino-rare-decays},
\ref{table-sgoldstino-flavor-decays} and 
\ref{table-sgoldstino-flavor-decays-2}), if corresponding processes
are allowed kinematically. 

\begin{figure}[htb]
\begin{center}
\begin{picture}(0,0)%
\epsfig{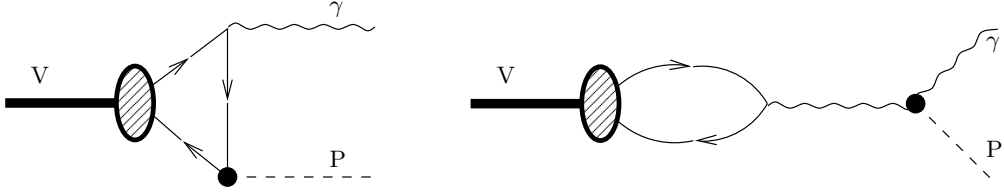}%
\end{picture}%
\setlength{\unitlength}{3947sp}%
\begingroup\makeatletter\ifx\SetFigFont\undefined%
\gdef\SetFigFont#1#2#3#4#5{%
  \reset@font\fontsize{#1}{#2pt}%
  \fontfamily{#3}\fontseries{#4}\fontshape{#5}%
  \selectfont}%
\fi\endgroup%
\begin{picture}(6300,1171)(156,-722)
\put(2251,-586){\makebox(0,0)[lb]{\smash{\SetFigFont{9}{10.8}{\rmdefault}{\mddefault}{\updefault}P}}}
\put(2251,389){\makebox(0,0)[lb]{\smash{\SetFigFont{9}{10.8}{\rmdefault}{\mddefault}{\updefault}$\gamma$}}}
\put(6376,-526){\makebox(0,0)[lb]{\smash{\SetFigFont{9}{10.8}{\rmdefault}{\mddefault}{\updefault}P}}}
\put(6376,164){\makebox(0,0)[lb]{\smash{\SetFigFont{9}{10.8}{\rmdefault}{\mddefault}{\updefault}$\gamma$}}}
\put(3301, -61){\makebox(0,0)[lb]{\smash{\SetFigFont{9}{10.8}{\rmdefault}{\mddefault}{\updefault}V}}}
\put(376,-61){\makebox(0,0)[lb]{\smash{\SetFigFont{9}{10.8}{\rmdefault}{\mddefault}{\updefault}V}}}
\end{picture}
\caption{Diagrams contributing to vector meson decay into sgoldstino
and photon.}
\label{vector-fig}
\end{center}
\end{figure}

We begin with constraints
independent of flavor violating terms in squark (slepton)
mass matrix. 
One obtains a set of limits on supersymmetry breaking scale by
considering Wilczek mechanism~\cite{wilczek} --- 
decay of neutral vector meson $V_{Q\bar{Q}}$ ($1^-$ state) 
into photon and (pseudo)scalar $S(P)$. 
There are two types of contributions into this process (see
Fig.~\ref{vector-fig}). 
The first one is
emission of real photons and (pseudo)scalars by quarks, while the
second is decay of virtual photons, emitted by quarks, 
into photons and (pseudo)scalars. The first process is governed by
fermion-sgoldstino coupling, while the second one emerges due to
interaction with a pair of photons. 
The relevant candidates on the role of $V_{Q\bar{Q}}$ are $J/\psi$, 
$\Upsilon$ and $\rho$-, $\omega$-, $\phi$-mesons.

Let us first consider heavy mesons, which may be described as
quasistationary systems. With account of 
effective lagrangian~(\ref{eef})
we obtain
\begin{equation}
{\Gamma(V_{Q\bar{Q}}\to S(P)\gamma)\over\Gamma(V_{Q\bar{Q}}\to\gamma\to
e^+e^-)}={M_V^2(A_Q\mp M_{\gamma\gamma})^2\over16\pi\alpha F^2}\;,
\label{wil-ratio}
\end{equation}
where $-(+)$ refers to decay into $S(P)$. 
We should compare the rate $\Gamma(V_{Q\bar{Q}}\to S(P)\gamma)$ with
current data on the rates $\Gamma(V_{Q\bar{Q}}\to\gamma+missing~energy)$, 
$\Gamma(V_{Q\bar{Q}}\to3\gamma)$ or 
$\Gamma(V_{Q\bar{Q}}\to\gamma+pair(s)~of~leptons(light~mesons))$ 
depending on $m_{S(P)}$
and superpartner mass spectrum (see discussion of sgoldstino decay
modes in section~\ref{sect2}). For illustration 
we set $M_{\gamma\gamma}=-A_Q=100$~GeV in our quantitative 
estimates, so vector
mesons would decay only into scalar sgoldstino. Eq.~(\ref{wil-ratio})
shows that the corresponding constraints on $\sqrt{F}$ scale as a square
root of the absolute value of the difference (sum) 
of $A_Q$ and $M_{\gamma\gamma}$, if one considers decay into scalar 
(pseudoscalar).   
\begin{table}[htb]
$$
\begin{array}{|c|c|c|c|}
\hline {\rm Experimental~limit} & X & m_S 
& \sqrt{F}{\rm Br}^{-1/4}_{(S\to X)}\\
\hline {\rm Br}(J/\psi\to S\gamma(S\to X))<5.5\cdot 10^{-5}~\cite{partridge} &
\gamma\gamma &<M_{J/\psi} & >180~{\rm GeV} \\ 
\hline
{\rm Br}(\Upsilon(1S)\to S\gamma(S\to X))<3.1\cdot
10^{-4}~\cite{albrecht} &\gamma\gamma&<0.1~{\rm GeV} &
>170~{\rm GeV} \\ 
\hline
{\rm Br}(\Upsilon(1S)\to S\gamma(S\to X))<3.1\cdot
10^{-4}~\cite{albrecht} &e^+e^-&<1.5~{\rm GeV} &
>170~{\rm GeV} \\
\hline
{\rm Br}(\Upsilon(1S)\to S\gamma(S\to X))<4\cdot
10^{-4}~\cite{albrecht} &\mu^+\mu^-,K^+K^-&<1.5~{\rm GeV} &
>160~{\rm GeV} \\ 
\hline
{\rm Br}(\Upsilon(1S)\to S\gamma(S\to X))<4\cdot
10^{-4}~\cite{albrecht} &\pi^+\pi^-&<1.5~{\rm GeV} &
>160~{\rm GeV} \\
\hline
{\rm Br}(\Upsilon(1S)\to S\gamma(S\to X))=(6.3\pm1.8)\cdot
10^{-5}~\cite{anastassov} &\pi^0\pi^0&>1.0~{\rm GeV} &
>440~{\rm GeV} \\
\hline
{\rm
Br}(\Upsilon(1S)\to S\gamma(S\to X))=(2\pm2)\cdot10^{-5}~\cite{fulton}

&2K^+2K^-&<M_{J/\psi}& >330~{\rm GeV} \\ 
\hline
\end{array}
$$
\caption{Constraints from decays of vector mesons on SUSY models
with light sgoldstinos decaying inside detector.}
\label{table-sgoldstino-rare-decays}
\end{table}
\begin{table}[htb]
$$
\begin{array}{|c|c|c|}
\hline {\rm Experimental~limit} & m_S 
& \sqrt{F}\;,{\rm GeV}\\
\hline 
{\rm Br}(J/\psi\to S\gamma)<1.4\cdot
10^{-5}~\cite{edwards} & \ll M_{J/\psi} & >260 \\ 
\hline
{\rm Br}(\Upsilon(1S)\to S\gamma)<1.3\cdot
10^{-5}~\cite{balest} &<5~{\rm GeV} & >370 \\ 
\hline
\end{array}
$$
\caption{Constraints from decays of vector mesons on SUSY models
with light sgoldstinos flying away from detector.}
\label{table-sgoldstino-rare-decays-out}
\end{table}

It turns out that constraints on $F$ from $\Upsilon$ 
decay into photons (leptons or light mesons), summarized in
Table~\ref{table-sgoldstino-rare-decays}, are of the same order as
limits from
processes with single photon and missing energy (see
Table~\ref{table-sgoldstino-rare-decays-out}) if corresponding
branching ratios for sgoldstino decay are roughly of
order one. The first type of
constraints (Table~\ref{table-sgoldstino-rare-decays}) 
is relevant for sgoldstino decaying within detector
(which is the case for $m_{S(P)}\gtrsim10$~MeV if 
$M_{\gamma\gamma}=A=100$~GeV);
these constraints scale as quartic root of the corresponding 
sgoldstino branching ratios. The second type of bounds
(Table~\ref{table-sgoldstino-rare-decays-out}) applies to 
lighter sgoldstino flying away from detector. We present in
Table~\ref{table-sgoldstino-rare-decays} only strongest
constraints on $\sqrt{F}$. 
Besides these, there is a number of other $\Upsilon$ decay
modes providing somewhat weaker constraints on  $F$:  
$\gamma\pi^+\pi^-K^+K^-$, 
$\gamma2\pi^+2\pi^-$, $\gamma3\pi^+3\pi^-$,
$\gamma2\pi^+2\pi^-K^+K^-$. 

One can show that limits on $\sqrt{F}$ 
from decays of light vector mesons ($\rho$, $\omega$, $\phi$) 
are weaker at least by an order of magnitude. 

Decays of $\Upsilon$ seem to have the best
sensitivity to flavor-conserving sgoldstino couplings if
$M_\Upsilon\gtrsim m_{S(P)}\gtrsim~a~few$~MeV. 
Since in the most part of the parameter space 
sgoldstino decays predominantly into 
two photons or two mesons, the most promising $\Upsilon$
decays are into three photons and into a photon and a pair of mesons. In
models with large trilinear soft terms, leptonic widths of sgoldstinos
become larger, and these modes become also interesting. 

\subsubsection{Flavor violating rare decays}
There is another type of processes to be considered. 
These are decays of charged particles: 
leptons or pseudoscalar mesons. The rates of these processes
are more model dependent because they are governed by flavor violating
soft terms. 

While the bounds on $\sqrt{F}$ coming from decays of leptons are the
same irrespectively of whether scalar or pseudoscalar sgoldstino is
created in the final state, in the flavor-violating hadronic processes
the creation of scalar sgoldstino is more important than the creation
of pseudoscalar sgoldstino (if they have similar masses). When we
discuss hadronic processes in what follows, we consider the emission of
scalar sgoldstino only. 
The simplest example is kaon decay $K^+\to\pi^+S$. In chiral theory
kaon conversion into pion is described by matrix element 
\begin{equation}
\la\pi^+|\bar{s}\gamma_\mu d|K^+\ra=\l f_+(k_K+k_\pi)_\mu+
f_-(k_K-k_\pi)_\mu\r\;,
\nonumber
\end{equation}
where $f_+=1$ and $f_-=0$ in the case of exact $SU(3)$ flavor
invariance. Then 
\begin{equation}
-\la\pi^+|\bar{s}d|K^+\ra=f_+\frac{m_K^2-m_\pi^2}{m_d+m_s}+
f_-\frac{m_S^2}{m_d+m_s}
\nonumber
\end{equation}
and in what follows we neglect $f_-$ contribution. 

In principle, there are two mechanisms of the decay of charged
particles with sgoldstinos in the final state. The first one 
is due to flavor-conserving sgoldstino 
interactions (with fermions and intermediate W-boson). 
The second one is due to flavor-changing terms in the 
low-energy effective interactions of light
sgoldstinos (see Eqs.(\ref{eef}), 
(\ref{llrrmix}); for instance, decay $K^+\to\pi^+S$ is due to 
${\cal L}_{eff}=-\frac{\tilde{m}_{D_{12}}^{LR~2}}{\sqrt{2}F}S\bar{s}d$). 
Hence, the second contribution emerges because of 
flavor violating interactions with fermions originating from
off-diagonal insertions in squark(slepton) mass matrix. 

As regards the first mechanism, it gives rise to constraints on $\sqrt{F}$
at the level of 100-250~GeV. We do not present these constraints 
explicitly, as they are at the same level or weaker than those summarized in 
Tables~\ref{table-sgoldstino-rare-decays}, 
\ref{table-sgoldstino-rare-decays-out}. 

The second mechanism is more interesting. The
corresponding constraints are presented in
Tables~\ref{table-sgoldstino-flavor-decays}, 
\ref{table-sgoldstino-flavor-decays-2}, 
\begin{table}[htb]
$$
\begin{array}{|c|c|c|c|}
\hline {\rm Experimental~limit} &
m_{S(P)}&(\delta_{ij})_{LR}& \sqrt{F}\;,{\rm GeV}\\
\hline {\rm Br}(\mu\to eS(P))<2.6\cdot10^{-6}~\cite{jodidio} & 
<m_\mu&\delta^l_{12}=1.7\cdot10^{-6} & >3\cdot10^4 \\ 
\hline
{\rm Br}(K^+\to\pi^+S)<3\cdot 10^{-10}~\cite{9708031} &
\simeq 0 &\delta^d_{12}=2.7\cdot10^{-3}  & >3.7\cdot 10^7 \\ 
\hline 
{\rm Br}(K^+\to\pi^+S)<5.2\cdot 10^{-10}~\cite{adler96} &
<80~{\rm MeV}&\delta^d_{12}=2.7\cdot10^{-3}  & >3.3\cdot10^7 \\ 
\hline
{\rm Br}(K^+\to\pi^+S)<10^{-8}~\cite{atiya} & 
\simeq180\div240~{\rm MeV}&\delta^d_{12}=2.7\cdot10^{-3}   & >1.6\cdot10^7 \\ 
\hline
\end{array}
$$
\caption{Constraints on SUSY models with light sgoldstinos flying away
from detector; bounds come 
from flavor changing decays of charged particles, if they are
allowed kinematically; we set 
flavor violating terms 
$\delta_{12}^d=\tilde{m}^{LR 2}_{D_{12}}/\tilde{m}^2_Q$, 
$\delta_{12}^l=\tilde{m}^{LR 2}_{L_{12}}/\tilde{m}^2_L$ 
to be equal to their current
bounds~\cite{masiero} at equal masses of squark and gluino, 
$M_3=\tilde{m}_Q$=500~GeV and equal masses of slepton and photino,
$\tilde{m}_l=M_{\tilde{\gamma}}$=100~GeV.}
\label{table-sgoldstino-flavor-decays}
\end{table}
\begin{table}[htb]
$$
\begin{array}{|c|c|c|c|c|}
\hline {\rm Experimental~limit} &X& m_{S(P)}&(\delta_{ij})_{LR} 
& \sqrt{F}{\rm Br}^{-1/4}_{S\to X}>\\[10pt]
\hline
{\rm Br}(\mu\to eS(P)(S(P)\to X)<7.2^{-11}~\cite{bolton88} & \gamma\gamma &
&\delta^l_{12}=1.7\cdot10^{-6} & 3.4\cdot10^5~{\rm GeV} \\ 
\hline
{\rm Br}(\mu\to eS(P)(S(P)\to X)<10^{-12}~\cite{bellgardt88} & e^+e^- &
>2m_e&\delta^l_{12}=1.7\cdot10^{-6} & 1.2\cdot10^5~{\rm GeV} \\ 
\hline
{\rm Br}(K^+\to\pi^+S(S\to X))<5\cdot
10^{-8}~\cite{kitching} 
&\gamma\gamma&
<100~{\rm MeV} & \delta^d_{12}=2.7\cdot10^{-3}&1.0\cdot10^7~{\rm GeV} \\ 
\hline
{\rm Br}(K^+\to\pi^+S(S\to X))<1.1\cdot 10^{-8}~\cite{Campagnari} &e^+e^-&
\simeq150\div340~{\rm MeV} &\delta^d_{12}=2.7\cdot10^{-3}& 1.5\cdot10^7~{\rm GeV} \\ 
\hline
{\rm Br}(K^+\to\pi^+S(S\to X))=(7.6\pm2.1)\cdot 10^{-8}~\cite{pdg} &\mu^+\mu^-&
>2m_\mu &\delta^d_{12}=2.7\cdot10^{-3}& 1.3\cdot10^7~{\rm GeV} \\ 
\hline
{\rm Br}(D^+\to\pi^+S(S\to X))<5.2\cdot10^{-5}~\cite{aitala} &e^+e^- &
>2m_e&\delta^u_{12}=3.1\cdot10^{-2} & 5.2\cdot10^5~{\rm GeV} \\ 
\hline
{\rm Br}(D^+\to\pi^+S(S\to X))<1.5\cdot10^{-5}~\cite{aitala} & \mu^+\mu^- &
>2m_\mu&\delta^u_{12}=3.1\cdot10^{-2} & 7\cdot10^5~{\rm GeV} \\ 
\hline
{\rm Br}(D^+\to\pi^+S(S\to X))=(2.2\pm0.4)\cdot10^{-3}~\cite{aitala} & \pi^+\pi^-&
>2m_\pi& \delta^u_{12}=3.1\cdot10^{-2}& 3.1\cdot10^5~{\rm GeV} \\ 
\hline
{\rm Br}(D_s^+\to K^+S(S\to X))<1.4\cdot10^{-4}~\cite{aitala} &\mu^+\mu^- &
>2m_\mu &\delta^u_{12}=3.1\cdot10^{-2} & 3.4\cdot10^5~{\rm GeV} \\ 
\hline
{\rm Br}(D_s^+\to K^+S(S\to X))<6\cdot10^{-4}~\cite{frabetti95f} &K^+K^- &
>2m_K&\delta^u_{12}=3.1\cdot10^{-2} & 2.4\cdot10^5~{\rm GeV} \\ 
\hline
{\rm Br}(B^+\to K^+S(S\to X))<6\cdot10^{-5}~\cite{avery89b} &e^+e^- &
>2m_e&\delta^d_{23}=1.6\cdot10^{-2} & 4.8\cdot10^5~{\rm GeV} \\ 
\hline
{\rm Br}(B^+\to K^+S(S\to X))<5.2\cdot10^{-6}~\cite{affolder} &\mu^+\mu^- &
>2m_\mu&\delta^d_{23}=1.6\cdot10^{-2} & 9.0\cdot10^5~{\rm GeV} \\ 
\hline
{\rm Br}(B^+\to K^+S(S\to X))<2.8\cdot10^{-5}~\cite{bergfeld96b} & \pi^+\pi^-&
>2m_\pi & \delta^d_{23}=1.6\cdot10^{-2} & 5.9\cdot10^5~{\rm GeV} \\ 
\hline
{\rm Br}(B^+\to\pi^+S(S\to X))<7.5\cdot10^{-5}~\cite{bergfeld96b} & K^+K^-&
>2m_K&\delta^d_{13}=3.3\cdot10^{-2}  & 6.6\cdot10^5~{\rm GeV} \\ 
\hline
{\rm Br}(B^+\to\pi^+S(S\to X))<4.1\cdot10^{-5}~\cite{bergfeld96b} & \pi^+\pi^-&
>2m_\pi&\delta^d_{13}=3.3\cdot10^{-2}  & 7.7\cdot10^5~{\rm GeV} \\ 
\hline
\end{array}
$$
\caption{Constraints on SUSY models with light sgoldstinos decaying
within detector, 
from search for flavor changing decays of charged particles; 
flavor violating terms $(\delta_{ij})_{LR}$ are the same as in 
Table~\ref{table-sgoldstino-flavor-decays}.}
\label{table-sgoldstino-flavor-decays-2}
\end{table}
where for definiteness we take flavor
violating off-diagonal 
insertions in squark(slepton) mass matrix to be equal to their 
current experimental limits~\cite{masiero} at 
$\tilde{m}_{squark}=M_3=500$~GeV, 
$\tilde{m}_{slepton}=100$~GeV. The limits on $\sqrt{F}$ scale as
inverted quartic root of bounds on meson branchings and as square
root of the off-diagonal elements $m_{ij}^{LR~2}$ 
in squark squared mass matrix; they depend crucially on the strength of
flavor violation in MSSM. Since hadronic and photonic modes 
usually dominate, limits on $\sqrt{F}$ coming from meson decays with a
pair of leptons in the final state (say, $K^+\to\pi^+S(S\to e^+e^-)$)
are weaker, but not more than by one or two orders of magnitude, as
compared to photonic and mesonic modes. 
Note, that similar constraints from
three-body decays of neutral mesons (like $B^0\to
K^0S(S\to\mu^+\mu^-)$) depend on the same coupling constants and
are generally weaker than limits from rare decays of charged mesons.  

From bounds presented in this section we conclude that sgoldstino
interactions may give large contributions into flavor changing 
rare decays, including
those forbidden in SM. In particular, in the case $F=1$~TeV$^2$, the 
constraints from 
processes with final light sgoldstino significantly strengthen 
the bounds on off-diagonal elements in squark and slepton mass matrices in
comparison with models where sgoldstinos decouple at low energies. 

Our analysis suggests that contributions of intermediate (virtual)
sgoldstinos into FCNC and lepton flavor violating processes may be
also significant. For instance, pseudoscalar mesons may decay through
light sgoldstino exchange. Also, there are
potentially important contributions to loop processes like
$K^0-\bar{K}^0$, $B^0-\bar{B}^0$ mixings, etc. These issues will be
considered elsewhere.

\subsection{Processes with two sgoldstinos}

Processes with two final sgoldstinos appear due to the presence
of two-sgoldstino interactions in low-energy effective lagrangian,
Eq.~(\ref{llrrmix}), and due to the double contribution of
one-sgoldstino interaction,
Eq.~(\ref{eef}). 
Of course, the corresponding amplitudes are highly
suppressed (by additional $F^{-1}$). 
Nevertheless, some of these processes are sensitive enough to place
constraints on $\sqrt{F}$ at the level of 1~TeV. 

Recall that two-sgoldstino coupling
constants~(\ref{llrrmix}) differ from one-sgoldstino
constants~(\ref{eef}). Indeed, they are
proportional to $LL$ and $RR$ insertions in scalar squared mass
matrix, while one-sgoldstino coupling constants are proportional to
$LR$ insertions. Note in this regard, that the current limits on
flavor changing squark masses $m_{ij}^{LL~2}$ and $m_{ij}^{RR~2}$ are
weaker than limits on $m_{ij}^{LR~2}$. 
Hence, it makes sense to consider processes where
two-sgoldstino couplings could be observed. 

Complete analysis of low-energy processes with two sgoldstinos
may be carried out along the same lines as for processes with
one sgoldstino. Instead of going through the limits systematically, we
discuss here only some examples in order to get the feeling of
sensitivity to $\sqrt{F}$. 

\subsubsection{Light neutral mesons}
\begin{figure}[htb]
\begin{center}
\begin{picture}(0,0)%
\epsfig{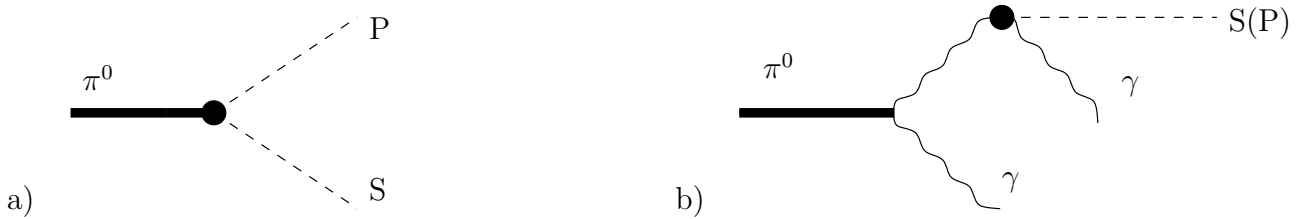}%
\end{picture}%
\setlength{\unitlength}{3947sp}%
\begingroup\makeatletter\ifx\SetFigFont\undefined%
\gdef\SetFigFont#1#2#3#4#5{%
  \reset@font\fontsize{#1}{#2pt}%
  \fontfamily{#3}\fontseries{#4}\fontshape{#5}%
  \selectfont}%
\fi\endgroup%
\begin{picture}(7812,1636)(601,-1037)
\put(1276,-211){\makebox(0,0)[lb]{\smash{\SetFigFont{12}{14.4}{\rmdefault}{\mddefault}{\updefault}$\pi^0$}}}
\put(3076,114){\makebox(0,0)[lb]{\smash{\SetFigFont{12}{14.4}{\rmdefault}{\mddefault}{\updefault}P}}}
\put(3076,-901){\makebox(0,0)[lb]{\smash{\SetFigFont{12}{14.4}{\rmdefault}{\mddefault}{\updefault}S}}}
\put(5551,-136){\makebox(0,0)[lb]{\smash{\SetFigFont{12}{14.4}{\rmdefault}{\mddefault}{\updefault}$\pi^0$}}}
\put(7051,-811){\makebox(0,0)[lb]{\smash{\SetFigFont{12}{14.4}{\rmdefault}{\mddefault}{\updefault}$\gamma$}}}
\put(7801,-211){\makebox(0,0)[lb]{\smash{\SetFigFont{12}{14.4}{\rmdefault}{\mddefault}{\updefault}$\gamma$}}}
\put(8476,164){\makebox(0,0)[lb]{\smash{\SetFigFont{12}{14.4}{\rmdefault}{\mddefault}{\updefault}S(P)}}}
\put(801,-961){\makebox(0,0)[lb]{\smash{\SetFigFont{12}{14.4}{\rmdefault}{\mddefault}{\updefault}a)}}}
\put(5001,-961){\makebox(0,0)[lb]{\smash{\SetFigFont{12}{14.4}{\rmdefault}{\mddefault}{\updefault}b)}}}
\end{picture}
\label{pion-fig}
\caption{a) The diagram illustrating 
$\pi^0$ decay into two sgoldstinos due to 
two-sgoldstino interaction; b) diagram of $\pi^0$ decay into 
two photons and sgoldstino due to one-sgoldstino interaction.}
\end{center}
\end{figure}

We begin with pion decay into two light sgoldstinos (see
Fig.~\ref{pion-fig}a). The relevant part of the effective lagrangian
reads 
\begin{equation}
\nonumber
{\cal L}=\frac{1}{4F^2}\l S\d_\mu P-P\d_\mu
 S\r\cdot
\l\l\tilde{m}_{U_{11}}^{LL~2}+\tilde{m}_{U_{11}}^{RR~2}\r
\bar{u}\gamma^\mu\gamma^5u+
\l\tilde{m}_{D_{11}}^{LL~2}+\tilde{m}_{D_{11}}^{RR~2}
\r\bar{d}\gamma^\mu\gamma^5 d\r\;.
\end{equation}
Then by making use of Eqs.~(\ref{pion-decay-constant-1}), 
(\ref{pion-decay-constant-2}) we obtain 
\begin{equation}
\Gamma(\pi^0\to SP)=\frac{f_\pi^2}{m_\pi}
\frac{[\tilde{m}_{U_{11}}^{LL~2}\!\!+\!\tilde{m}_{U_{11}}^{RR~2}\!\!-\!
\tilde{m}_{D_{11}}^{LL~2}\!\!-\!\tilde{m}_{D_{11}}^{RR~2}]^2}{128\pi F^2}
\frac{\l m_S^2-m_P^2\r^2}{F^2}
\sqrt{\l1\!+\!\frac{m_P^2-m_S^2}{m_\pi^2}\r^2\!\!-\!4{m_P^2\over m_\pi^2}}\;. 
\label{dyr}
\end{equation}
This rate is proportional to $(m_S^2-m_P^2)$, so, as expected, 
it vanishes in the massless limit, $m_S\;,m_P\to 0$. 
In order to examine the
sensitivity of this process, 
let us neglect the phase volume dependence and take 
$|m_S^2-m_P^2|\sim m_\pi^2/4$. 
If we set the value in the square bracket equal to  $2\tilde{m}_Q^2$ 
and choose $\tilde{m}_Q=500$~GeV, 
we obtain the limits presented in 
Table~\ref{table-two-sgoldstino-flavor-decays}. A few remarks are in
order. First, these bounds 
may be irrelevant in some theories 
because $\sqrt{F}$ should not be significantly smaller than any of the
soft terms. Second, the constraint from pion disappearance (i.e., 
from Br($\pi^0\to SP$)) is valid only if sgoldstinos fly away from
detector. For $m_{S(P)}\simeq m_\pi/2$ this is the case if
$M_{\gamma\gamma},A_e<10$~GeV, which is not forbidden by current experiments. 
Third, these limits are obtained at tuned sgoldstino masses and, in
general, they are weaker (see Eq.~(\ref{dyr})). 
\begin{table}[htb]
$$
\begin{array}{|c|c|c|}
\hline {\rm Experimental~limit} & 
\sqrt{F}\;,{\rm GeV}\\
\hline 
{\rm Br}(\pi^0\to SP)<8.3\cdot 10^{-7}~\cite{nnatiya} &
 >150 \\
\hline {\rm Br}(\pi^0\to SP(S\to2\gamma\;,P\to2\gamma)<2\cdot 
10^{-8}~\cite{mcdonough} &
 >240\cdot{\rm Br}^{-1/8}_{S\to\gamma\gamma}
{\rm Br}^{-1/8}_{P\to\gamma\gamma} \\ 
\hline 
\end{array}
$$
\caption{Constraints on SUSY models
with light sgoldstinos from neutral pion decay due to 
two-sgoldstino coupling to matter fields; these constraints are evaluated
at $|m_S^2-m_P^2|=m_\pi^2/4$ and $\tilde{m}_Q=500$~GeV (see text).}
\label{table-two-sgoldstino-flavor-decays}
\end{table}

These results do not depend on flavor-violating couplings and are 
of the same order of magnitude as the limits presented in
Tables~\ref{table-sgoldstino-rare-decays-out},
\ref{table-sgoldstino-rare-decays}. However, the limits presented in
Table~\ref{table-two-sgoldstino-flavor-decays} scale as inverted octopic root
of the corresponding pion partial width 
(see Eq.~(\ref{dyr})). 

To illustrate that two-sgoldstino processes 
may impose more
stringent constraints than one-sgoldstino processes with the same content
of final SM particles, let us estimate
the one-sgoldstino contribution to pion decay into four
photons. Namely let us consider 
emission of sgoldstinos from the photon
legs of pion (see Figure~\ref{pion-fig}b). If sgoldstino decays within
detector into photons, this would correspond to
four-photon decay of pion. Pion-photon anomalous amplitude reads
$$
A(\pi\to\gamma\gamma)=
-\frac{\alpha}{\pi f_\pi}\epsilon_{\mu\nu\rho\sigma}
\epsilon^\mu_1\epsilon^\nu_2q^\rho_3q^\sigma_2\;,
$$
where $q_1$, $q_2$ are the photon momenta. Then the 
corresponding squared matrix element of $\pi^0\to\gamma\gamma S(P)$ is
$$
|M|^2=4\frac{\alpha^2}{\pi^2}
\frac{M_{\gamma\gamma}^2f_\pi^2}{F^2}\l (q_1p)^2+(q_1q_3)^2\r\;,
$$
where $p$ and $q_1$, $q_3$ are momenta of sgoldstino and outgoing photons,
respectively. Neglecting sgoldstino mass we estimate the decay width as 
\begin{equation}
\Gamma(\pi^0\to\gamma\gamma S(P))=
{1\over 32}\frac{\alpha^2m_\pi^3}{f_\pi^2\pi^4}
\frac{M_{\gamma\gamma}^2m_\pi^2}{F^2}\;.
\label{pion-gamma-gamma-S}
\end{equation}
One can check that Eq.~(\ref{pion-gamma-gamma-S}) gives weaker bound 
on $\sqrt{F}$ than the limit presented in the second row of 
Table~\ref{table-two-sgoldstino-flavor-decays} if
$M_{\gamma\gamma}=100$~GeV and $\tilde{m}_Q=500$~GeV. 

Let us now evaluate the bounds from decays of neutral kaons 
due to two-sgoldstino flavor violating couplings. The
effective lagrangian reads
\begin{eqnarray}
{\cal L}=\frac{1}{4F^2}\l S\d_\mu P-P\d_\mu
 S\r\cdot
\biggl(\l\tilde{m}_{D_{21}}^{LL~2}+\tilde{m}_{D_{21}}^{RR~2}
\r\bar{s}\gamma^\mu\gamma^5 d+
\l\tilde{m}_{D_{21}}^{LL~2}-\tilde{m}_{D_{21}}^{RR~2}
\r\bar{s}\gamma^\mu d\nonumber
\\+\l\tilde{m}_{D_{12}}^{LL~2}+\tilde{m}_{D_{12}}^{RR~2}
\r\bar{d}\gamma^\mu\gamma^5 s+
\l\tilde{m}_{D_{12}}^{LL~2}-\tilde{m}_{D_{12}}^{RR~2}
\r\bar{d}\gamma^\mu s\biggr)\;.
\nonumber
\end{eqnarray}
One can show, that only the measurements of branching ratios 
of $K^0_L$ impose interesting 
constraints on $F$, whereas current limits on rare $K^0_S$ decays
provide weak constraints on $\sqrt{F}$. We obtain 
by making use of chiral theory 
\begin{equation}
\Gamma(K^0_L\to SP)=\frac{f_K^2}{m_K}
\frac{[\tilde{m}_{D_{21}}^{LL~2}\!+\tilde{m}_{D_{12}}^{LL~2}\!+
\tilde{m}_{D_{21}}^{RR~2}\!+\tilde{m}_{D_{12}}^{RR~2}]^2}{512\pi F^2}
\frac{\l m_S^2-m_P^2\r^2}{F^2}
\sqrt{\l1\!+\frac{m_P^2-m_S^2}{m_K^2}\r^2\!\!-4{m_P^2\over m_K^2}}\;. 
\nonumber
\end{equation}
Note that in the limit of CP conservation, $LL$ and $RR$ 
squark mass matrices are real and symmetric, and the sum in the bracket equal
to $2\l\tilde{m}_{D_{21}}^{LL~2}\!+\tilde{m}_{D_{21}}^{RR~2}\r$.

In analogy to the discussion of pion decays, 
let us neglect the phase volume dependence and set 
$|m_S^2-m_P^2|\simeq m_K^2/4$, $f_K=160$~MeV. 
If we set the sum in the bracket equal
to $4{\rm Re}~\tilde{m}_{D_{21}}^{LL~2}$ 
and impose on ${\rm Re}~\tilde{m}_{D_{21}}^{LL~2}$ 
current constraints from the absence
of FCNC~\cite{masiero} at squark mass $\tilde{m}_Q=500$~GeV, we obtain the
limits presented in 
Table~\ref{table-two-sgoldstino-flavor-decays-2}. 
\begin{table}[htb]
$$
\begin{array}{|c|c|}
\hline {\rm Experimental~limit} & \sqrt{F}\;,~{\rm
GeV} \\
\hline 
{\rm Br}(K^0_L\to SP\to e^+e^-\gamma\gamma)=
(6.9\pm1.0)\cdot 10^{-7}~\cite{pdg} 
& >1.9\cdot10^3\cdot{\rm Br}^{-1/8}_{S(P)\to e^+e^-}
{\rm Br}^{-1/8}_{P(S)\to\gamma\gamma}
\\ 
\hline 
{\rm Br}(K^0_L\to SP\to e^+e^-e^+e^-)=(4.1\pm0.8)\cdot
10^{-8}~\cite{pdg} &
>2.7\cdot10^3\cdot{\rm Br}^{-1/8}_{S(P)\to e^+e^-}
{\rm Br}^{-1/8}_{P(S)\to e^+e^-}
\\ 
\hline 
{\rm Br}(K^0_L\to SP\to\mu^+\mu^-e^+e^-)=(2.9^{+6.7}_{-2.4})\cdot
10^{-9}~\cite{gu} & 
>3\cdot10^3\cdot{\rm Br}^{-1/8}_{S(P)\to\mu^+\mu^-}
{\rm Br}^{-1/8}_{P(S)\to e^+e^-}
\\ 
\hline
{\rm Br}(K^0_L\to SP\to e^+e^-\pi^+\pi^-)=(3.5\pm0.6)\cdot
10^{-7}~\cite{pdg} &
>2.1\cdot10^3\cdot{\rm Br}^{-1/8}_{S(P)\to \pi^+\pi^-}
{\rm Br}^{-1/8}_{P(S)\to e^+e^-}
\\ 
\hline
\end{array}
$$
\caption{Constraints on SUSY models
with light sgoldstinos coming from decays of $K_L^0$ due to 
two-sgoldstino flavor-violating coupling to matter fields; we set 
real parts of the flavor violating term, 
$(\delta_{12})_{LL}=\tilde{m}^{LL~2}_{D_{12}}/\tilde{m}^2_Q$, 
equal to its current bound, 
Re~$(\delta_{12})_{LL}=4.6\cdot10^{-2}$~\cite{masiero} 
at equal masses of squarks and gluino, 
$M_3=\tilde{m}_Q$=500~GeV; these constraints are evaluated
at $|m_S^2-m_P^2|=m_{K^0}^2/4$.}
\label{table-two-sgoldstino-flavor-decays-2}
\end{table}
Note that limits
from kaon decays into a leptonic pair and a pair of mesons(photons) are
usually more significant than limits from decays into four leptons,
because of small sgoldstino decay branching ratio into leptons (see
Figure~\ref{Jfig}). These bounds on
$\sqrt{F}$ are obtained at tuned sgoldstino masses, $|m_S^2-m_P^2|\simeq
m_K^2/4$, and generally the bounds are somewhat weaker. 

We are not aware of limits on decays $K_L^0\to4\gamma$ and 
$K_L^0\to\pi^+\pi^-\gamma\gamma$. If it would be possible to measure
(or limit) their branching ratios at the level of $10^{-7}$, the
sensitivity of experiments to two-sgoldstino couplings would increase,
because the photonic decay usually dominates over leptonic decay of
sgoldstinos. 

\subsubsection{Decays of heavy mesons}
In analogy with light mesons 
we consider now heavy neutral mesons living 
sufficiently long, $D^0$, $B^0$ and $B^0_s$. 
We make use of the approach similar to the 
chiral theory in order to describe
their interaction with sgoldstinos; in the following we set 
$f_{B^0_s}=f_{B^0}=f_{D^0}$=200~MeV. 
The limits obtained with the same assumptions as above 
about sgoldstino masses and values of flavor violating couplings are
listed in 
Table~\ref{table-two-sgoldstino-meson-decays}. All remarks concerning these
assumptions given in previous subsection may be repeated here. 
\begin{table}[htb]
$$
\begin{array}{|c|c|c|c|}
\hline {\rm Experimental~limit} & |m_S^2-m_P^2| 
&(\delta_{ij})_{LL}&  \sqrt{F}>\;,{\rm GeV}\\
\hline  
{\rm Br}(D^0\to SP\to2\pi^+2\pi^-)=(7.3\pm0.5)\cdot 10^{-3}~\cite{pdg} &
\simeq m^2_{D^0}/4&\delta^u_{12}=1.0\cdot10^{-1} & 360\cdot{\rm Br}^{-1/8}_{S\to \pi^+\pi^-}
{\rm Br}^{-1/8}_{P\to \pi^+\pi^-} \\ 
\hline 
{\rm Br}(D^0\to SP\to K^+K^-\pi^+\pi^-)<8\cdot 10^{-4}~\cite{frabetti} &
\simeq m^2_{D^0}/4&\delta^u_{12}=1.0\cdot10^{-1} & 340\cdot{\rm Br}^{-1/8}_{S\to \pi^+\pi^-}
{\rm Br}^{-1/8}_{P\to \pi^+\pi^-} \\ 
\hline
{\rm Br}(D^0\to SP\to3\pi^+3\pi^-)=(4\pm3)\cdot 10^{-4}~\cite{barlag-65}&
\simeq m^2_{D^0}/4 &\delta^u_{12}=1.0\cdot10^{-1}& 
390\cdot{\rm Br}^{-1/8}_{S(P)\to 2\pi^+2\pi^-}
{\rm Br}^{-1/8}_{P(S)\to \pi^+\pi^-} \\ 
\hline 
{\rm Br}(B^0\to SP\to 2\pi^+2\pi^-)<2.3\cdot 
10^{-4}~\cite{adam} &
\simeq m^2_{B^0}/4 &\delta^d_{13}=9.8\cdot10^{-2}& 710\cdot{\rm Br}^{-1/8}_{S\to \pi^+\pi^-}
{\rm Br}^{-1/8}_{P\to \pi^+\pi^-} \\ 
\hline
\end{array}
$$
\caption{Constraints on SUSY models
with light sgoldstinos from decays of heavy neutral mesons due to 
two-sgoldstino coupling to matter fields.}
\label{table-two-sgoldstino-meson-decays}
\end{table}
It turns out that constraints on two-sgoldstino coupling 
from current bounds on $B_s$ decay modes are rather weak.  

To conclude this subsection, we note that 
``two-sgoldstino'' bounds scale as inverted octopic root of
meson branching ratios, so eight order improvements (!) in measurements of
corresponding partial widths are required to strengthen the bounds on
$\sqrt{F}$ by an order of magnitude. 

\section{Discussions and Conclusions}

We have considered models with light ($m_{S(P)}<a~few$~GeV) superpartners of
goldstino. The constraints on their effective couplings to SM particles
have been presented. By making use of these constraints we have estimated
the sensitivity of low energy experiments to the 
scale of supersymmetry breaking and gravitino mass. 

Let us compare our results with constraints coming from
high energy processes~\cite{dicus,0001025,0005076}. 
If we ignore sgoldstino, then 
in models with light gravitino current direct bound on supersymmetry
breaking scale is
obtained from Tevatron, $\sqrt{F}<$217~GeV~\cite{tevatron}. 
The upgraded Tevatron may be able
to cover the range of $\sqrt{F}$ up to 
$\sqrt{F}\simeq$290~GeV~\cite{0006162}, while LHC will be capable to
reach $\sqrt{F}<$1.6~TeV~\cite{9801329}. In models with light sgoldstinos, collider
experiments become more sensitive to the scale of sypersymmetry breaking. 
Most powerful among the operating machines, LEP and
Tevatron, give a constraint of order 1~TeV on supersymmetry breaking
scale in models with light sgoldstinos. Indeed, it was found 
in Ref.~\cite{0001025} that with the LEP luminosity of 100~pb$^{-1}$, at
$\sqrt{F}=1\div1.5$~TeV one $e^+e^-\to SZ(\gamma)$ event would
occur, and ten $e^+e^-\to e^+e^-S$ events would appear at
$\sqrt{F}=1.5$~TeV. At Tevatron, about 10 events in
$pp\to S\gamma(Z)$ channel, and $10^5$ events in $pp\to S$
channel would be produced at $\sqrt{F}=1$~TeV~\cite{0005076}. 
This gives rise to a possibility to detect
sgoldstino, if it decays inside detector into photons and $\sqrt{F}$
is not larger than $1.5\div2$~TeV. However, these numbers have been obtained
in a model with heavier superpartner scale, and, hence, with larger
sgoldstino couplings, than we assumed in our paper. 
For that set of parameters bounds on $\sqrt{F}$
derived in our paper from processes originating due to 
flavor-conserved sgoldstino couplings 
should be stronger at least by a factor of 1.5~. 

One important remark concerns the sensitivity of collider experiments to light
particles. The currently available analyses carried out by
experimental collaborations are relevant only for 
fairly heavy sgoldstino ($M\gtrsim20$~GeV). In this paper we have discussed lighter
particles;  in this sense our results may be considered as 
complementary to those derived up to now from LEP and Tevatron
experiments.  

From constraints presented in this paper we conclude that the
sgoldstino signal is not likely to be observed at LEP and Tevatron 
if sgoldstinos are lighter than a few GeV 
and flavor-violating processes in MSSM are not
extremely suppressed. 
One observes that precision measurements at low
energies are promising for confirming directly such a
model. The astrophysical bounds are usually stronger than
laboratory ones, though they become invalid for $m_S$ and $m_P$ larger than a
few MeV. 

Among the laboratory processes, the most sensitive to very light sgoldstinos
are propagation of laser beam in magnetic field and reactor
experiments. For heavier sgoldstinos 
the most sensitive processes are $\Upsilon$ decays for
flavor-conserving sgoldstino couplings and charged kaon decays 
for flavor-violating sgoldstino couplings. 

{\bf Acknowledgments}

The author is indebted to F.~Bezrukov, P.~Onyisi, 
A.~Ovchinnikov, A.~Penin, 
A.~Pivovarov and V.~Rubakov for useful discussions. 
This work was supported in part by RFBR 
grant 99-01-18410, CRDF grant 6603, Swiss Science Foundation,
grant 7SUPJ062239, Russian Academy of Science, JRP
grant \# 37 and by ISSEP fellowship. 

%%%%%%%%%%%%%%%%%%%%%%%%%%%%%%%%%%%%%%%%%%%%%%%%%%%%%%%%%
\def\ijmp#1#2#3{{\it Int. Jour. Mod. Phys. }{\bf #1~} (19#2) #3}
\def\pl#1#2#3{{\it Phys. Lett. }{\bf B#1~} (19#2) #3}
\def\zp#1#2#3{{\it Z. Phys. }{\bf C#1~} (19#2) #3}
\def\prl#1#2#3{{\it Phys. Rev. Lett. }{\bf #1~} (19#2) #3}
\def\rmp#1#2#3{{\it Rev. Mod. Phys. }{\bf #1~} (19#2) #3}
\def\prep#1#2#3{{\it Phys. Rep. }{\bf #1~} (19#2) #3}
\def\pr#1#2#3{{\it Phys. Rev. }{\bf D#1~} (19#2) #3}
\def\np#1#2#3{{\it Nucl. Phys. }{\bf B#1~} (19#2) #3}
\def\mpl#1#2#3{{\it Mod. Phys. Lett. }{\bf A#1~} (19#2) #3}
\def\arnps#1#2#3{{\it Annu. Rev. Nucl. Part. Sci. }{\bf #1~} (19#2) #3}
\def\sjnp#1#2#3{{\it Sov. J. Nucl. Phys. }{\bf #1~} (19#2) #3}
\def\jetp#1#2#3{{\it JETP Lett. }{\bf #1~} (19#2) #3}
\def\app#1#2#3{{\it Acta Phys. Polon. }{\bf #1~} (19#2) #3}
\def\rnc#1#2#3{{\it Riv. Nuovo Cim. }{\bf #1~} (19#2) #3}
\def\ap#1#2#3{{\it Ann. Phys. }{\bf #1~} (19#2) #3}
\def\ptp#1#2#3{{\it Prog. Theor. Phys. }{\bf #1~} (19#2) #3}
\def\spu#1#2#3{{\it Sov. Phys. Usp.}{\bf #1~} (19#2) #3}
\def\apj#1#2#3{{\it Ap. J.}{\bf #1~} (19#2) #3}
\def\epj#1#2#3{{\it Eur.\ Phys.\ J. }{\bf C#1~} (19#2) #3}
\def\pu#1#2#3{{\it Phys.-Usp. }{\bf #1~} (19#2) #3}
\def\nc#1#2#3{{\it Nuovo Cim. }{\bf A#1~} (19#2) #3}
%%%%%%%%%%%%%%%%%%%%%%%%%%%%%%%%%%%%%%%%%%%%%%%%%%%%%%%%%


{\small
\begin{thebibliography}{99}					
\bibitem{ellis} J.~Ellis, K.~Enqvist, D.~Nanopoulos, \pl{147}{84}{99};
J.~Ellis, K.~Enqvist, D.~Nanopoulos, \pl{151}{85}{357}. 
\bibitem{no-scale} T.~Bhattacharya, P.~Roy, \pr{38}{88}{2284}. 
\bibitem{gmm} G.~Giudice, R.~Rattazzi, , \prep{322}{99}{419}; 
   S.~Dubovsky, D.~Gorbunov, S.~Troitsky, \pu{42}{99}{705}.
\bibitem{naturalness} A.~Brignole, F.~Feruglio, F.~Zwirner,
\pl{438}{98}{89}. 
\bibitem{bhat} T.~Bhattacharya, P.~Roy, \pl{206}{88}{655}.
\bibitem{9612253} J.~Grifols, R.~Mohapatra, A.~Riotto,
\pl{400}{97}{124}. 
\bibitem{cosmo-paper} T.~Gherghetta, \pl{423}{98}{311}.
\bibitem{dicus} D.~Dicus, S.~Nandi, J.~Woodside, \pr{41}{90}{2347}; 
D.~Dicus, S.~Nandi, \pr{56}{97}{4166}.
\bibitem{0001025} E.~Perazzi, G.~Ridolfi, F.~Zwirner, 
{\it Nucl. Phys. }{\bf B574} (2000) 3.
\bibitem{0005076} E.~Perazzi, G.~Ridolfi, F.~Zwirner, {\it Signatures
of massive sgoldstinos at hadron colliders}, hep-ph/0005076.
\bibitem{9904367} A.~Brignole, E.~Perazzi, F.~Zwirner, JHEP 9909
(1999) 002. 
\bibitem{masiero} F. Gabbiani {\em et al.}, \np{477}{96}{321}; 
M. Ciuchini {\em et al.}, {\it JHEP}{\bf 9810} (1998) 008.
\bibitem{gold-flavor} A.~Brignole, A.~Rossi, {\it Flavour
non-conservation in goldstino interactions}, hep-ph/0006036.
\bibitem{tevatron} CDF Collaboration (T. Affolder, {\em et al.}), 
{\it Phys. Rev. Lett.}~{\bf 85}~(2000) 1378. 
\bibitem{voloshin-zakharov} M.~Voloshin and V.~Zakharov,
\prl{45}{80}{688}. 
\bibitem{novikov} D.~Gross, S.~Treiman, F.~Wilczek, \pr{19}{79}{2188}; 
V.~Novikov {\em et al.}, \np{165}{80}{55}.
\bibitem{pich} A.~Pich, {\it Effective Field Theory} 
(Lectures at the 1997 Les Houches Summer School "Probing the Standard
       Model of Particle Interactions"), hep-ph/9806303.
\bibitem{HelScope} S.~Moriyama {\em et al.}, \pl{434}{98}{147}. 
\bibitem{Solar-crystal} SOLAX Collaboration (F.~Avignone {\em et
al.}), \prl{81}{98}{5068};
R.~J.~Creswick {\em et al.}, \pl{427}{98}{235}.
\bibitem{SN} J.~Brockway, E.~Carlson, G.~Raffelt, \pl{383}{96}{439}.
\bibitem{HBS} G.~Raffelt, {\it Stars as Laboratories for Fundamental
Physics} (University of Chicago Press, Chicago, 1996).
\bibitem{Back} E.~Masso, R.~Toldra, \pr{52}{97}{7967}.
\bibitem{RedGiant} G.~Raffelt and A.~Weiss, \pr{51}{95}{1495}.
\bibitem{HBS-eP} D.~Dicus {\em et al.}, \pr{18}{78}{1829}.
\bibitem{HBS-eS} J.~Grifols, E.~Masso, \pl{173}{86}{237}; J.~Grifols,
E.~Masso, S.~Peris, \mpl{4}{89}{311}.
\bibitem{R} G.~Raffelt, {\it Stars as Laboratories for Fundamental
Physics} (Univ. of Chicago Press, Chicago, 1996). 
\bibitem{cameron} R.~Cameron {\em et al.}, \pr{47}{93}{3707}.
\bibitem{NOMAD} NOMAD Collaboration, (P. Astier {\em et al.}), 
{\it Phys. Lett. }{\bf B479} (2000) 371. 
\bibitem{altmann} M.~Altmann {\em et al.}, \zp{68}{95}{221}.
\bibitem{koch} H.~Koch, O.~Schult, \nc{96}{86}{182}.
\bibitem{reactor} T.~Donnelly {\em et al.}, \pr{18}{78}{1607}. 
\bibitem{wilczek} F.~Wilczek, \prl{39}{77}{1304}.
\bibitem{partridge} R.~Partridge {\em et al.}, \prl{44}{80}{712}.
\bibitem{albrecht} ARGUS Collaboration (H. Albrecht {\em et al.}),
\pl{179}{86}{403}. 
\bibitem{anastassov} CLEO Collaboration (A. Anastassov {\em et al.}),
\prl{82}{99}{286}.  
\bibitem{fulton} R. Fulton {\em et al.}, \pr{41}{90}{1401}.
\bibitem{edwards} C.~Edwards {\em et al.}, \prl{48}{82}{903}.
\bibitem{balest} CLEO Collaboration (R. Balest {\em et al.}), 
\pr{51}{95}{2053}.   
\bibitem{jodidio} A. Jodidio {\em et al.}, \pr{34}{86}{1967}, {\it 
Erratum} - \pr{37}{88}{237}. 
\bibitem{9708031} E787 Collaboration (S.~Adler {\em et al.}), 
\prl{79}{97}{2204}.
\bibitem{adler96} S.~Adler {\em et al.}, \prl{76}{96}{1421}. 
\bibitem{atiya} M.~Atiya {\em et al.}, \prl{64}{90}{21}.
\bibitem{bolton88} R.~Bolton {\em et al.}, \pr{38}{88}{2077}. 
\bibitem{bellgardt88} SINDRUM Collaboration (U. Bellgardt {\em et
al.}), \np{299}{88}{1}.
\bibitem{kitching} E787 Collaboration (P.~Kitching {\em et al.}), 
\prl{79}{97}{4079}. 
\bibitem{Campagnari} C.~Alliegro {\em et al.}, \prl{68}{92}{278}.
\bibitem{pdg} Particle Data Group, {\it Eur.\ Phys.\ J. }{\bf C15} (2000) 1.
\bibitem{aitala} E791 Collaboration (E.M. Aitala {\em et al.}),
\pl{462}{99}{401}. 
\bibitem{frabetti95f} E687 Collaboration (P.L. Frabetti {\em et al.}),
\pl{363}{95}{259}.
\bibitem{avery89b} P. Avery {\em et al.}, \pl{223}{89}{470}.
\bibitem{affolder} CDF Collaboration (T. Affolder {\em et al.}),
\prl{83}{99}{3378}. 
\bibitem{bergfeld96b} CLEO Collaboration (T. Bergfeld {\em et al.}), 
\prl{77}{96}{4503}.
\bibitem{nnatiya} M.~Atiya {\em et al.}, \prl{66}{91}{2189}.
\bibitem{mcdonough} J.~McDonough {\em et al.}, \pr{38}{88}{2121}. 
\bibitem{gu} P. Gu {\em et al.}, \prl{76}{96}{4312}. 
\bibitem{frabetti} E687 Collaboration (P.L. Frabetti {\em et al.}), 
\pl{354}{95}{486}.
\bibitem{barlag-65}  ACCMOR Collaboration (S.~Barlag {\em et al.}), \zp{55}{92}{383}.
\bibitem{adam} DELPHI Collaboration (W.~Adam {\em et al.}), \zp{72}{96}{207}.
\bibitem{0006162} S. Ambrosanio {\em et al.}, {\it Report of the Beyond the MSSM Subgroup for the Tevatron Run II
SUSY/Higgs Workshop}, hep-ph/0006162.
\bibitem{9801329} A. Brignole {\em et al.}, \np{526}{98}{136}. 
\end{thebibliography}}
\end{document}